\documentclass[12pt]{iopart}

\usepackage[utf8]{inputenc}
\usepackage{cite}
\usepackage[normalem]{ulem}

\usepackage{xparse}
\usepackage{etoolbox}

\newbool{PrintVersion}
\ifdefined\PrintVersion
    \booltrue{PrintVersion}
\fi

\usepackage[english]{babel}
\usepackage[babel=true]{csquotes} 

\usepackage{amssymb}

\usepackage{bm}
\usepackage{braket}
\usepackage{upgreek}

\usepackage[babel]{microtype}

\usepackage{graphicx}
\usepackage[hyperref, svgnames]{xcolor}
\usepackage{tabularx}
\usepackage{booktabs}

\usepackage{hyperref}

\hypersetup
{
    \ifbool{PrintVersion}{hidelinks}{colorlinks},
    linkcolor={red!60!black},
    citecolor={blue!60!black},
    urlcolor={green!40!black},
    unicode=true,
    bookmarksnumbered=true,
    pdfstartview=FitH,
    pdfpagemode=UseNone,
    pdfpagelayout=TwoColumnRight,
    pdfdisplaydoctitle=true,
    pdfduplex=DuplexFlipLongEdge,
}

\usepackage[noabbrev,capitalize,nameinlink]{cleveref} 
\crefalias{subequation}{equation}

\usepackage{acronym}

\newcommand{\euler}{\mathrm{e}}
\newcommand{\imag}{\mathrm{i}}
\newcommand{\RE}{\mathrm{Re}}
\newcommand{\IM}{\mathrm{Im}}
\newcommand{\overset}[2]{\ensuremath{\mathop{\kern\z@\mbox{#2}}\limits^{\mbox{\scriptsize #1}}}}
\newcommand{\SI}[2]{{#1}\, \mathrm{#2}}

\begin{document}
\title[GPR-based method for the localization of exceptional points]{Gaussian-process-regression-based method for the localization of exceptional points in complex resonance spectra}
\author{Patrick Egenlauf, Patric Rommel and J\"org Main}
\address{Institut f\"ur Theoretische Physik 1, Universit\"at Stuttgart, 70550 Stuttgart, Germany}
\ead{main@itp1.uni-stuttgart.de}

\date{\today}

\acrodef{gpr}[GPR]{Gaussian process regression}
\acrodef{gp}[GP]{Gaussian process}
\acrodefplural{gp}[GPs]{Gaussian processes}
\acrodef{rbf}[RBF]{radial basis function}
\acrodef{ep}[EP]{exceptional point}
\acrodef{dp}[DP]{diabolical point}
\acrodef{lml}[LML]{log marginal likelihood}
\acrodef{cu2o}[Cu$_2$O]{cuprous oxide}

\begin{abstract}
Resonances in open quantum systems depending on at least two
controllable parameters can show the phenomenon of exceptional points
(EPs), where not only the eigenvalues but also the eigenvectors of two
or more resonances coalesce.
Their exact localization in the parameter space is challenging, in
particular in systems, where the computation of the quantum spectra and
resonances is numerically very expensive.
We introduce an efficient machine learning algorithm to find
exceptional points based on Gaussian process regression (GPR).
The GPR-model is trained with an initial set of eigenvalue pairs
belonging to an EP and used for a first estimation of the EP position
via a numerically cheap root search.
The estimate is then improved iteratively by adding selected exact
eigenvalue pairs as training points to the GPR-model.
The GPR-based method is developed and tested on a simple
low-dimensional matrix model and then applied to a challenging real
physical system, viz., the localization of EPs in the resonance
spectra of excitons in cuprous oxide in external electric and magnetic
fields.
The precise computation of EPs, by taking into account the complete
valence band structure and central-cell corrections of the crystal,
can be the basis for the experimental observation of EPs in this
system.
\end{abstract}

\section{Introduction}
\label{sec:intro}
In open quantum systems, resonances can occur~\cite{Moiseyev1998,Ho1983}.
These are quasi-bound states which can decay.
By introducing a complex scaling, e.g.\ according to Reinhardt
\cite{Reinhardt1982}, and thus non-Hermitian operators, the complex
energy eigenvalues of the resonances can be calculated.
Here, the real part represents their energy, while the imaginary part
unveils their lifetime.
Depening on the studied problem, the computational effort required to
calculate the eigenvalues and eigenvectors can be enormous.

In a parameter-dependent Hamiltonian, certain choices of the
parameters can lead to the degeneration of two or more
resonances.
A special case of this is the so-called \ac{ep}, at which not only the
eigenvalues but also the eigenstates degenerate\cite{Kato1966}.
Thus, 
$n\ge 2$ resonances coalesce in the general case of an
EP$n$.
\acp{ep} have attracted much attention during recent years.
They appear in PT-symmetric quantum systems~\cite{Bender1998,Bender1999},
photonic systems~\cite{Miri2019,Ozdemir2019,Ergoktas2022},
electronic circuits~\cite{Stehmann2004},
atoms in external fields~\cite{Cartarius2007,Cartarius2009},
Bose-Einstein condensates~\cite{Gutoehrlein2016},
and have been applied for enhanced sensing~\cite{Hodaei2017,Wiersig2020}.
A property of the most prominent case of an
EP{2} is that the two associated eigenvalues exchange their
positions after one adiabatic orbit in parameter space around the \ac{ep}.
In experiments, external fields often can be used as conveniently
controllable parameters for the manipulation of resonance spectra.
In 2007 the existence of these \acp{ep} was proven numerically for the
hydrogen atom in electric and magnetic fields~\cite{Cartarius2007},
however, at field strengths much higher than laboratory fields.
Due to limitations especially in magnetic field strengths, \acp{ep} in
the hydrogen atom have not yet been observed experimentally.

In 2014, a remarkable discovery by Kazimierczuk \textit{et al.}
\cite{Kazimierczuk2014} revealed a hydrogen-like spectrum up to a
principal quantum number of $n=25$ within \ac{cu2o}. These spectral
features are caused by excitons, quasi-particles in a semiconductor
consisting of electron and hole. To a first approximation, they
closely resemble their atomic counterpart, the hydrogen atom. However,
the fact that the excitons are environed by \ac{cu2o} necessitates
consideration of the details of the valence band structure to
precisely describe the observed spectrum.

This experiment opened up research into giant Rydberg excitons.
While the achieved principal quantum numbers are far lower than those
reached in atomic systems, the modified material parameters in the
crystal still lead to the strongly magnified properties known from
Rydberg atoms.
Similarly, for \ac{cu2o} the field strengths needed to observe
\acp{ep} of resonances with small quantum numbers are much lower
compared to the field strengths needed in the hydrogen atom, which is
why it is more favorable to find \acp{ep} in this system.

Under suitable conditions, the search for \acp{ep} can be
  reduced to a root search problem for the distance between the eigenvalues
  forming the \ac{ep}.
Existing methods to find \acp{ep} are
  the three-point~\cite{Uzdin2010} and octagon
  method~\cite{Feldmaier2016}. They are based on a Taylor expansion
of the spacing between the eigenvalues in the local vicinity
  of the \ac{ep} up to linear or quadratic order in the
  parameters, respectively.
They work similarly to a Newton algorithm, iteratively generating improved estimates of the \ac{ep} by finding the roots of these linear or quadratic approximations.
They thus may fail if the starting points are not in close proximity to the \ac{ep}. 
The numerical search for \acp{ep} in reference~\cite{Feldmaier2016}
has been applied to a hydrogen-like model of excitons in external
fields, however, for a detailed comparison of experimental and
theoretical results in cuprous oxide, the above mentioned band
structure terms need to be considered.
This increases the computational cost drastically for each
diagonalization of the Hamiltonian due to its complexity, and thus
methods using a large number of such diagonalizations cannot be applied.
Hence, an improved method is required to accurately and efficiently
identify \acp{ep} in complex systems like \ac{cu2o} with external
fields.

Inspired by the remarkable advances in machine learning, especially
within the realm of physics, a novel method on the foundation of
\ac{gpr} \cite{rasmussen_gaussian_2005} is developed. As a prominent
member of the supervised machine learning family, \ac{gpr} serves as a powerful and innovative approach to predict the positions of
\acp{ep}.
The used data to train a \ac{gpr} model is obtained by
simulations. Hence, the error is only due to numerical
inaccuracies. Unlike neural networks, \ac{gpr} offers the advantage
that the provided training points can be passed through almost
  exactly to a numerical accuracy, which is a key motivation for
its utilization. Also, unlike kernel ridge regression which is
  deterministic \cite{Saunders1998}, the \ac{gpr} model provides a
  prediction and also a corresponding variance at chosen points,
  which are both used in our application of the method.
The basic idea is to perform the root search over the cheaply
  calculated output of a \ac{gpr} model, instead of using the
  expensive exact diagonalizations or low-precision linear or
  quadratic Taylor approximations.
Yet, the optimization of the searching process goes beyond the usage of machine learning.
An efficient algorithm is devised to enhance the search for \acp{ep}
in \ac{cu2o}, which contributes to the discovery of promising \acp{ep}
candidates and thus enables a possible experimental verification of
these data.
As a preparatory step, we introduce the stepwise grouping
  algorithm to detect the presence of \acp{ep} via their
  characteristic exchange behavior \cite{Heiss1991} in a dataset of
  parameter-dependent eigenvalue spectra, and filter out the
  associated eigenvalues. The obtained data is then used as an initial
  training set for the \ac{gpr} model, which learns the relationship
  between parameter values and eigenvalues.
After the root search over the model output, which produces a first
estimate of the \ac{ep} position, we calculate the exact eigenvalues
at the predicted parameters of the \ac{ep}.
An improved model is then obtained by appending the additional new
eigenvalue pair to the training data.
This is repeated until the desired precision is achieved.
The procedure is more efficient than existing methods, since the
model keeps the information from the initial data set and from
previous iteration steps and can thus approximate the eigenvalue distance
very well after only very few iterations.
Furthermore, the iterative process can now start at rather large
distance from the true solution and needs fewer time-consuming exact
diagonalizations.

The paper is organized as follows.
In \cref{sec:Resonances} we give a general background to resonances
and \acp{ep} in quantum systems.
In \cref{subsec:GPRMethod} we introduce our GPR-based method for
localizing \acp{ep}, which is then applied in
\cref{sec:Results} to
excitons in \ac{cu2o} in parallel electric and magnetic fields.
We conclude with a summary and outlook in \cref{sec:Conclusion}.

\section{Methods and materials}
\label{sec:MethodsAndMaterials}

\subsection{Resonances and exceptional points in quantum mechanics}
\label{sec:Resonances}
Quantum states of real-world systems are rarely perfectly bound. External fields and coupling to the environment create decay channels, leading to quasibound states with limited lifetimes. These states are called resonances. Mathematically, they are conveniently described using complex energies $\tilde{E}$, where the imaginary part is proportional to the decay rate
\begin{equation}
\tilde{E} = E - \mathrm{i} \Gamma / 2\,,
\end{equation}
with the spectral energy position $E$ and the FWHM linewidth $\Gamma$. When applying the time evolution operator to a state with such a complex energy, the resulting probability density decays exponentially with the decay rate $\Gamma/\hbar$.

Since Hermitian operators only have real eigenvalues, these complex
energies can not be straightforwardly calculated as the eigenstates of
a standard quantum mechanical Hamiltonian, but can be computed with an
appropriate non-Hermitian modification of the Hamiltonian, e.g., by
application of the complex-coordinate-rotation
method~\cite{Reinhardt1982,Ho1983,moiseyev_2011}.

In non-Hermitian quantum mechanics the eigenstates of the Hamiltonian
are not necessarily orthogonal, and in case of degenerate states it is
even possible, that not only the eigenvalues but also the eigenvectors
of two or more resonances coalesce.
These degeneracies are called \acp{ep}~\cite{Kato1966,Heiss2000}.
In order to visualize an \ac{ep}2 --- where two resonances form an \ac{ep} --- and its behavior, a linear non-Hermitian map represented by the two-dimensional matrix \cite{Kato1966}
\begin{equation}
	\boldsymbol{M}(\kappa) = \left(\begin{array}{cc}
		1 & \kappa \\ 
		\kappa & -1
	\end{array}\right)
	\label{eq:EPSimpleExampleMatrix}
\end{equation}
can be used as a simple example with $\kappa \in \mathbb{C}$. Despite being a straightforward illustration, all significant aspects of \acp{ep} can be demonstrated since the close vicinity of an \ac{ep} in an $n$-dimensional system can be expressed by a two-dimensional problem \cite{Heiss2000}. 
\begin{figure}
	\centering
	\includegraphics[width=0.8\textwidth]{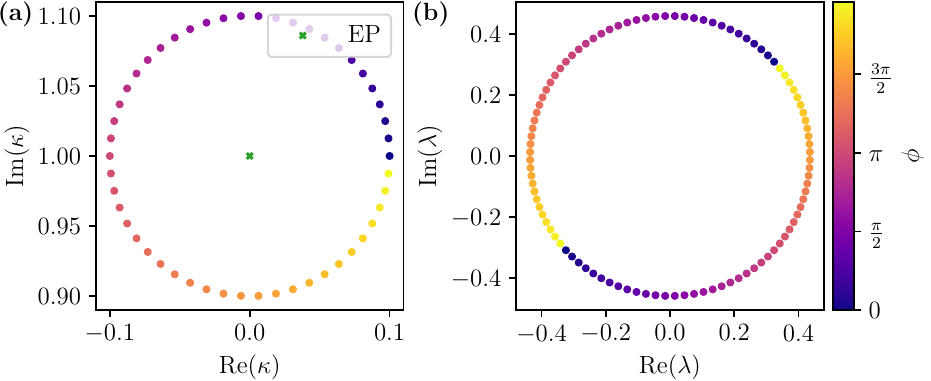}%
	\caption{\textbf{(a)} The \ac{ep} $\kappa_+ = \imag$ of the simple example \labelcref{eq:EPSimpleExampleMatrix} is encircled in the parameter space $\kappa$. Each point in the complex plane is described by the angle $\phi$ in Euler form as depicted in \cref{eq:EPSimpleExampleCircularParametrization}. \textbf{(b)} This leads to an exchange of the positions of the two eigenvalues in the complex energy plane. The eigenvalues calculated for each point on the circle are marked by the respective color of the color bar indicating the angle $\phi$, in order to illustrate the path of each eigenvalue and the associated permutation.} %
	\label{fig:epExample2d}
\end{figure}
Calculating the eigenvalues of \cref{eq:EPSimpleExampleMatrix} results in
\begin{equation}
	\lambda_1 = \sqrt{1 + \kappa^2}\,,~ \lambda_2 = -\sqrt{1 + \kappa^2}\,.
	\label{eq:EPSimpleExampleEigenvalues}
\end{equation}
It is evident that the eigenvalues are two branches of the same analytic function in $\kappa$. They coalesce at $\kappa_\pm = \pm \imag$, which means that there exist two \acp{ep} in this system. The corresponding eigenvectors are
\begin{equation}
	\boldsymbol{v}_1(\kappa) = \left(\begin{array}{c}- \kappa \\ 1 - \sqrt{1 + \kappa^2}\end{array}\right) \,,~ \boldsymbol{v}_2(\kappa) = \left(\begin{array}{c}- \kappa \\ 1 + \sqrt{1 + \kappa^2}\end{array}\right)
\end{equation}
from which it is obvious that these also coalesce at $\kappa_\pm = \pm\imag$ and that there is therefore only one linearly independent eigenvector
\begin{equation}
	\boldsymbol{v}(\pm\imag) = \left(\begin{array}{c}\mp\imag \\ 1\end{array}\right) \,.
\end{equation}
The existence of an \ac{ep} in the system~\eref{eq:EPSimpleExampleMatrix} 
can be revealed by a characteristic exchange behavior of the
eigenvalues in the complex plane when encircling the \ac{ep} in the
parameter space~\cite{Heiss1991}.
In the system~\eref{eq:EPSimpleExampleMatrix} the exchange behavior
can be visualized by parameterizing
\begin{equation}
\kappa(\phi) = \imag + \varrho\euler^{\imag\phi} \,,
\label{eq:EPSimpleExampleCircularParametrization}
\end{equation}
which describes a circle with radius $\varrho$ around the \ac{ep} $\kappa_+ = \imag$.
This orbit is shown in \cref{fig:epExample2d}a, where the color bar denotes the angle $\phi$. The permutation of the eigenvalues in the complex energy plane can be seen in \cref{fig:epExample2d}b. After one closed loop in the parameter space the eigenvalues exchange their positions. The path of each eigenvalue is depicted by the colors of the color bar. Due to the simplicity of this example, this permutation can also be shown analytically by substituting the circular parametrization \labelcref{eq:EPSimpleExampleCircularParametrization} into the eigenvalues from \cref{eq:EPSimpleExampleEigenvalues}
\begin{eqnarray}
\lambda_{1,2} = \pm \sqrt{1 + \left(\imag + \varrho \euler^{\imag\phi}\right)^2}& = \pm \sqrt{\varrho} \euler^{\imag\frac{\phi}{2}} \sqrt{2\imag + \varrho \euler^{\imag \phi}}
              \stackrel{\varrho \ll 2}{\approx} \pm \sqrt{2 \varrho} \euler^{\imag\frac{\pi}{4}} \euler^{\imag\frac{\phi}{2}} \,,
\end{eqnarray}
which leads to
\begin{equation}
\lambda_1 = \sqrt{2 \varrho} \euler^{\imag \left( \frac{\pi}{4} + \frac{\phi}{2}\right)} \quad \mathrm{and} \quad \lambda_2 = \sqrt{2 \varrho} \euler^{\imag \left( \frac{5\pi}{4} + \frac{\phi}{2}\right)} \,.
\end{equation}
After one closed loop in the parameter plane, i.e.\ $\phi = 2\pi$, $\lambda_1$ changes to $\lambda_2$ and vice versa, which is exactly the behavior shown in \cref{fig:epExample2d}.
Due to the small radius both paths of the eigenvalues in the complex energy plane build a half circle. For larger radii or elliptical orbits this shape can be much more complex.
\acp{ep} can occur in open physical quantum systems, depending on at
least a two-dimensional parameter space.
However, for a high-dimensional non-Hermitian Hamiltonian their
localization is a very nontrivial task.

To systemize the localization of \acp{ep}, we first reduce the
  task to a two-dimensional root search problem.
We start with a dataset of parameter-dependent eigenvalue spectra, in which the characteristic exchange behavior has been detected. We then consider the complex eigenvalues $\lambda_1$ and $\lambda_2$
  depending on the complex control parameter $\kappa$, which are associated with a specific
\ac{ep}, and introduce the analytical complex functions
\begin{equation}
	p \equiv \left(\lambda_1 - \lambda_2\right)^2 \label{eq:EvDiff}\,, 
\end{equation}
\begin{equation}
	s \equiv \frac{1}{2} \left(\lambda_1 + \lambda_2\right) \label{eq:EvSum}\,,
\end{equation}
with $p=0$ exactly at the \ac{ep}.
To ensure analyticity, the eigenvalue difference is squared in the
variable $p$, owing to the
square-root behavior of an EP2.
The \ac{ep} can now be localized via a root search for $p=0$.
For our iterative procedure, we also want to keep the information
given by the centroid $s$, since $p$ and $s$ together uniquely determine the eigenvalue pair,
and thus help select the correct eigenvalue
pair among all eigenvalues of a given diagonalization as described
below in \cref{sec:iteration}.

Existing methods for the localization of \acp{ep} include the three-point
method~\cite{Uzdin2010} and the octagon method~\cite{Feldmaier2016}.
Both methods use a Taylor expansion to approximate
$p$, with the difference that the three-point method only uses terms
up to linear order, while the octagon method also considers quadratic
terms.
Especially the three-point method converges towards the \ac{ep} only
if the starting points are in the close vicinity of the \ac{ep},
similar to Newton's method.
The main goal of this paper is to develop a method for the
localization of \acp{ep}, which, on the one hand is not restricted to
the local vicinity of the \ac{ep} but acts more globally in the
parameter space, and, on the other hand, requires a low number of
exact diagonalizations to calculate the eigenvalues, as this is the
most time-consuming and computationally expensive step.
To this aim we resort to \ac{gpr} in what follows.

\subsection{\ac{gpr}-based method to find \acp{ep}}
\label{subsec:GPRMethod}

\begin{figure}[t]
	\centering
	\includegraphics[width=0.8\textwidth]{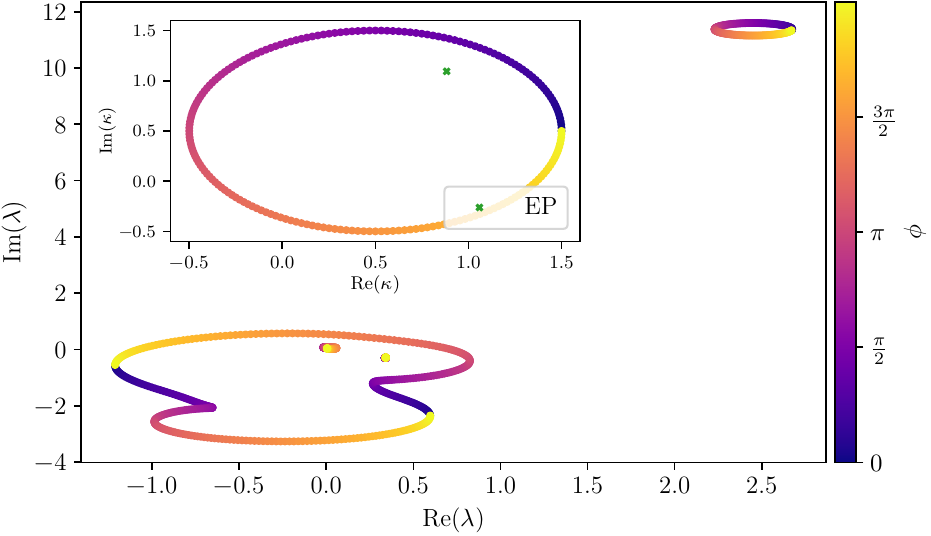}%
	\caption[Energy plane of two five-dimensional matrix models]{The investigated five-dimensional matrix model is visualized with its respective energy and parameter plane. The \ac{ep}, found via a two-dimensional root search on $p$ without a \ac{gpr} model, is marked as a green cross. Due to its dimensionality, five eigenvalues are visible, and their course can be traced via the color bar, which represents the angle $\phi$, when the \ac{ep} is encircled. The model shows one permutation, which indicates the existence of an \ac{ep}. The shape of this permutation is complex because of the large radius and the fact that the \ac{ep} is not at the center of the orbit.}%
	\label{fig:Toy5DExampleEnergyKappa}
\end{figure}

The \ac{gpr}-based method for the localization of \acp{ep} shall be
applicable to high-dimensional systems, which are much more
challenging than the simple model~\eref{eq:EPSimpleExampleMatrix}.
To address the complexity of more challenging systems, a
complex-symmetric five-dimensional matrix
\begin{equation}
  M(\kappa) = \left(\begin{array}{ccccc}
	1      & \kappa & M_{13} & M_{14} & M_{15}\\ 
	\kappa &    -1  & M_{23} & M_{24} & M_{25}\\
	M_{13} & M_{23} & M_{33} & M_{34} & M_{35}\\
	M_{14} & M_{24} & M_{34} & M_{44} & M_{45}\\
	M_{15} & M_{25} & M_{35} & M_{45} & M_{55}
\end{array}\right)
\label{eq:M5dim}
\end{equation}
is introduced, which depends on a complex parameter $\kappa$ and
complex matrix elements $M_{ij}$, $1 \leq i \leq 5$, $3 \leq j \leq
5$, $i \leq j$, where the real and imaginary parts are random numbers.
This matrix model, which has no analytic solution, is illustrated in
\cref{fig:Toy5DExampleEnergyKappa}.
The parameter plane with the orbit around the \ac{ep}, highlighted in
green, as well as the energy plane are presented.
Here, the computational
effort to diagonalize the matrix and hence get the eigenvalues is low,
which is why the \ac{ep} can be found by performing a two-dimensional
root search on $p$ without the \ac{gpr} model, where only the two
eigenvalues belonging to the \ac{ep} are considered. The result is
marked as a green cross in the parameter plane. Due to its
dimensionality, each diagonalization yields five eigenvalues. In the
model (cf.\ \cref{fig:Toy5DExampleEnergyKappa}), two eigenvalues lie
within the visible permutation of the two eigenvalues of the
\ac{ep}. All eigenvalues not belonging to an \ac{ep} form a closed
orbit similar to the one in parameter space.
Given that the size of the path encirling the \ac{ep} is relatively
large and the \ac{ep} is not localized in its center, the path of the
two permuting eigenvalues exhibit significantly greater complexity
compared to the relatively straightforward two-dimensional example
depicted in \cref{fig:epExample2d}.
As the \ac{gpr} model necessitates the
incorporation of the two eigenvalues associated with the \ac{ep}, the
observed permutation needs to be distinguished from the other
eigenvalues.

The \ac{gpr} method is first presented using the five-dimensional
matrix model to verify its functionality as well as its accuracy.
The method requires as a starting point a set of eigenvalue pairs
calculated along a closed path in the parameter space encircling the
\ac{ep}, which exhibit the characteristic permutation of the two
eigenvalues.
This is achieved with the stepwise grouping algorithm described in
\cref{sec:sga}.
The centroids and squared differences $s$ and $p$ (see \cref{eq:EvDiff,eq:EvSum}) of
these eigenvalue pairs are used as initial training set for a
\ac{gpr} model, which yields a first estimation of the \ac{ep}
position in parameter space via a two-dimensional numerical root
search.
This step is introduced in \cref{sec:gpr-model}.
The estimate of the \ac{ep} is then improved iteratively.
Therefore an exact diagonalization of the Hamiltonian is performed at
the estimated \ac{ep} position.
The eigenvalue pair belonging to the \ac{ep} is filtered out of the
complete set of eigenvalues and added to the training set for an
improved \ac{gpr} model, which yields an improved estimate of the
\ac{ep} position via a numerical root search.
The steps of this loop are discussed in \cref{sec:iteration}.
The iteration stops when one of the convergence criteria presented in
\cref{sec:convergence} is fulfilled.

\subsubsection{Stepwise grouping algorithm}
\label{sec:sga}
The starting point for searching for an \ac{ep} is a set of eigenvalue
spectra calculated along a closed path in the parameter space.
Along the closed path most eigenvalues start at and return to the same
point, except for those eigenvalue pairs belonging to an \ac{ep}.
In large data sets it is a nontrivial task to select these eigenvalue
pairs.
By application of the stepwise grouping algorithm we sort the
eigenvalues into groups in such a way that
each resulting group consists of the parameter-eigenvalue pairs
tracing a single parameter-dependent path through the eigenvalue plane.
An \ac{ep} is then characterized by a path, where the initial and
final points do not coincide, i.e., two eigenvalues exchange their
positions along that path.
Selecting such paths yields an initial set of eigenvalue pairs
belonging to an \ac{ep}.

In our application, the numerically calculated eigenvalues are initially
sorted by their real parts.
For a given parameter $\kappa_i$, let
$\boldsymbol{\lambda}(\kappa_i)$ denote the corresponding set of eigenvalues obtained from the diagonalization. In the limit where the $\{\kappa_i\}$ constitute a continous curve, $\boldsymbol{\lambda}(\{\kappa_i\})$ 
gives a set of continous paths $\{\lambda_j(\{\kappa_i\})\}$ through the eigenvalue space. In practice, we have a finite set of discrete values of $\kappa_i$. For the stepwise grouping algorithm, we suppose that this set, while finite and discrete, still covers a closed loop in the parameter space sufficiently densely that for every path $j_0$, the eigenvalue $\lambda_{j_0}(\kappa_{i+1})$ is closer according to some distance measure to  $\lambda_{j_0}(\kappa_i)$ than the eigenvalues  $\lambda_j(\kappa_{i+1})$ belonging to other paths $j \neq j_0$ are. Starting from a given point in the parameter space $\kappa_0$, the algorithm then follows the loop, in each step appending to each path the eigenvalue closest to the one from the previous step. Care must be taken if an eigenvalue would be the prefered continuation of multiple paths. In this case, the algorithm appends the value to the path where the distance is smallest and renews the sorting for the other eigenvalues at parameter $\kappa_{i+1}$.
To define a suitable distance measure between eigenvalues, multiple options exist. A simple option would be to use the Euclidean distance $d_\mathrm{e} = |\lambda_{m} - \lambda_{n} |$ in the complex plane, which is used for the matrix model.
A more sophisticated approach is utilized for the application to
cuprous oxide (see \cref{sec:Results}), which additionally uses
information contained in the eigenstates. We can form a vector
$\boldsymbol\Psi$ where we combine the complex eigenvalues with
additional admixtures of quantum numbers characterizing the state. To
obtain a distance measure, we can then use the cosine
similarity
\begin{equation}
	d_\mathrm{c} = 1 - \frac{(\boldsymbol\Psi_m , \boldsymbol\Psi_n)}{||\boldsymbol\Psi_m||\,||\boldsymbol\Psi_n||} = 1 - \frac{\sum_{l=1}^M (\Psi_l)_m (\Psi_l)_n}{\sqrt{\sum_{l=1}^M (\Psi_l)_m^2 \sum_{l=1}^M (\Psi_l)_n^2}} ,
	\end{equation}
where $M$ is the dimension of the vectors $\boldsymbol\Psi$. The
cosine similarity is subtracted from $1$ to obtain a measure that is
small when the vectors are similar.

Note that in order to achieve accurate sorting of resonances, a
sufficient number of points must be computed along the orbit to avoid
significant variations in the eigenvalues between consecutive points.
The exact number of points required depends on the radius of the orbit.
When applied to the five-dimensional matrix model in
\cref{fig:Toy5DExampleEnergyKappa} the algorithm selects the one pair
of eigenvalues that exchange along the closed path in the parameter
space.

\subsubsection{Model training}
\label{sec:gpr-model}
We want to interpolate numerically calculated values of the
  functions $p$ and $s$ defined in \cref{eq:EvDiff,eq:EvSum}, which
  are obtained via expensive diagonalizations.
    Since there is no experimental or other source of noise, we do not
    want to treat the training data as an approximation to some underlying
    function which is contaminated with small errors. Instead, we want
    to treat the training data as exact and interpolate to other input
    values. We are thus looking for a method that faithfully extends a
    set of functions from a known set of arguments to other input values.
    \ac{gpr}-based methods are a much better choice for these
    requirements than, e.g., neural networks, where predicted values
    at training points can deviate significantly from the given values.
    By construction, they treat the data as known
    observations in a Bayesian sense (modulo some uncertainty,
    $\sigma_n\sim 10^{-6}$ (see \cref{eq:Matern5_2,eq:GPRMeanFunction,eq:LogMarginalLikelihood} in \ref{sec:gprInML}), which is negligibly small for our application)
    and  output a conditional distribution of fit functions.
  Other methods would potentially also fulfill our requirements and
  would probably lead to similar results.
  For example, assuming a suitable choice of kernel and hyperparameters,
  the posterior mean function obtained by \ac{gpr} is the same as the
  fit obtained via kernel ridge regression~\cite{Wang2022}.
  \ac{gpr} also has the advantage of allowing hyperparameter optimization via the
  log marginal likelihood, instead of requiring a grid search, and it provides a variance for each prediction. This variance, which corresponds to the model uncertainty about the prediction, is incorporated in the method as outlined in \cref{sec:iteration}.
The theoretical background and implementation of the \ac{gpr} method
is described in \ref{sec:gprInML}. For our kernel, we use the Mat\'{e}rn
type class as given in \cref{eq:Matern5_2} with $\nu = 5/2$.
The other hyperparameters are optimized during the training step as
outlined in \ref{sec:GPRTraining}.

\begin{figure}%
	\centering%
	\includegraphics[width=0.9\textwidth]{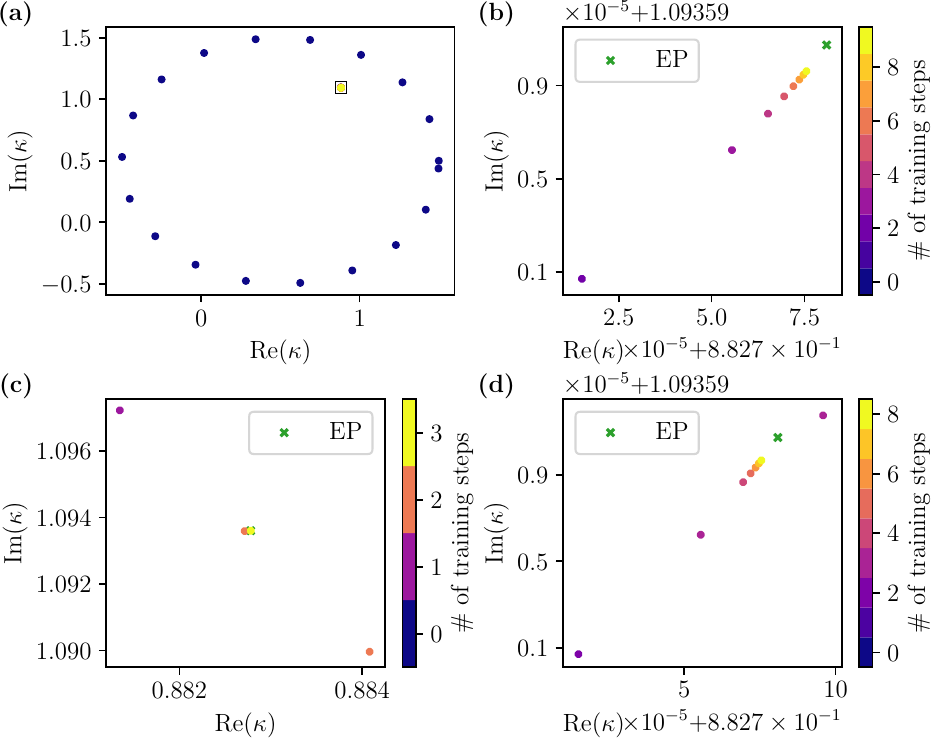}%
	\caption{Illustration of the iterative process by means of the five-dimensional matrix model. \textbf{(a)} The points on the orbit around the \ac{ep}, marked in blue, and their associated eigenvalues are used as initial training set. \textbf{(b)} The model slowly approaches the \ac{ep}, marked as a green cross. After nine training steps the \ac{gpr} model is converged and the Euclidean distance between the last model prediction and the exact \ac{ep} is $d_\mathrm{e} = 5.526\times 10^{-6}$. Two different attempts are made to optimize convergence and reduce the number of exact diagonalizations. \textbf{(c)} First, an additional training point is added after the second iteration to explore the energy plane. For this purpose, the difference of the last two predictions is calculated and added to the second prediction. This leads to convergence after the third training step, i.e.\ after the fourth diagonalization (considering the additional point). Not only the number of diagonalizations is significantly reduced, but also the Euclidean distance to $d_\mathrm{e} = 1.342\times 10^{-6}$. \textbf{(d)} Similarly to the previous approach, an additional training point is added after the third iteration. This does not reduce the number of diagonalizations (nine training steps, ten diagonalizations) nor does it improve convergence ($d_\mathrm{e} = 5.537\times 10^{-6}$).}%
	\label{fig:Toy5DIterativeProcessModel2}%
\end{figure}

The selected set of eigenvalue pairs are used to train two \ac{gpr}
models in a way that the two functions $p,s \in \mathbb{C}$, defined in
\cref{eq:EvDiff,eq:EvSum}, can be predicted for chosen values of
$\kappa$, as mappings $\mathbb{R}^2\to \mathbb{R}^2$, i.e.
\begin{equation}
	\boldsymbol{\kappa} = \left(\begin{array}{c}\RE(\kappa) \\ \IM(\kappa)\end{array}\right)
\to \boldsymbol{p} = \left(\begin{array}{c}\RE(p)
  \\ \IM(p)\end{array}\right) \mathrm{~and~}
\boldsymbol{s} = \left(\begin{array}{c}\RE(s)
  \\ \IM(s)\end{array}\right)\, .
\end{equation}
Due to the coalescence of the eigenvalues, $\boldsymbol{p}$ should be
zero at the \ac{ep}.
Performing a two-dimensional root search on the model output
$\boldsymbol{p}$ yields a prediction of
$\boldsymbol{\kappa}_\mathrm{EP}$ for the \ac{ep}.
The predictions of $\boldsymbol{s}$ will be used in a later step.
For the five-dimensional matrix model \eref{eq:M5dim}
the obtaind value of $\boldsymbol{\kappa}_\mathrm{EP}$ is already close
to the exact position of the \ac{ep}.
It is marked in \cref{fig:Toy5DIterativeProcessModel2}a (dots in the
small square window) and as a violet dot (first iteration step) in the
enlarged area in \cref{fig:Toy5DIterativeProcessModel2}c.

\subsubsection{Iterative improvement of the \ac{gpr}-model}
\label{sec:iteration}
To construct an iterative process, an exact diagonalization is
performed with the predicted parameter value.
However, the identification of the two eigenvalues
associated with the \ac{ep} among all obtained eigenvalues in each iteration is not a straightforward
task for higher-dimensional systems.
The diagonalization of a five-dimensional matrix yields a total set of
five eigenvalues, but only the ones related to the \ac{ep} warrant
selection.
In contrast to the problem of finding the eigenvalues belonging to the \ac{ep} among the initial eigenvalue set in \cref{sec:sga}, there is no permutation pattern to guide the selection process here.
To find the two eigenvalues associated with the \ac{ep} in each
iteration step a similarity measure is introduced
based on a Gaussian distribution
\begin{equation}
	\mathcal{N}(\mu, \sigma) = \euler^{-\frac{\left(x - \mu\right)^2}{2\sigma^2}}\,.
	\label{eq:Toy5DGaussianDistribution}
\end{equation}
Here, the value of $s$ in \cref{eq:EvSum} plays a significant
role. The model predictions $\boldsymbol{p}$ and $\boldsymbol{s}$ at
the predicted $\kappa$ point are considered as $\mu$ and compared to
the exact eigenvalues of the matrix at this $\kappa$ point.
The values $\sigma^2$ are obtained as the model uncertainties in the prediction
given by the \ac{gpr} method, see~\cref{eq:GPRCovarianceFunction} in
\ref{sec:GPRTraining}.
To perform this comparison, all possible pairs of eigenvalues are formed
(10 eigenvalue pairs in case of the five-dimensional matrix model
\eref{eq:M5dim}), and their respective $p$ and $s$ values are
calculated. They take the part of $x$ in 
\cref{eq:Toy5DGaussianDistribution}.
As the exponential function is monotonic, it is
sufficient to consider the exponent, yielding the pair-discrepancy
\begin{eqnarray}
	c &=& \frac{\left(\RE(p) - \RE(p_\mathrm{m})\right)^2}{2 \sigma_\mathrm{p,Re}^2} + \frac{\left(\IM(p) - \IM(p_\mathrm{m})\right)^2}{2 \sigma_\mathrm{p,Im}^2} \nonumber\\  
	&+& \frac{\left(\RE(s) - \RE(s_\mathrm{m})\right)^2}{2 \sigma_\mathrm{s,Re}^2} + \frac{\left(\IM(s) - \IM(s_\mathrm{m})\right)^2}{2 \sigma_\mathrm{s,Im}^2}\,,
	\label{eq:Toy5DDiscrepancy}
\end{eqnarray}
which is small when there is good agreement between model prediction
and exact eigenvalue pair. The index `m' indicates a model-predicted
value and $\sigma^2$ is the variance obtained from the \ac{gpr} model
for the respective prediction.
This pair-discrepancy is calculated for all possible eigenvalue pairs and the one with the lowest value is chosen.
Ideally, there should be a large gap between the smallest and second smallest $c$ value to make sure, that the correct eigenvalue pair is selected. 
\begin{figure}[t]
	\centering
	\includegraphics[width=0.8\textwidth]{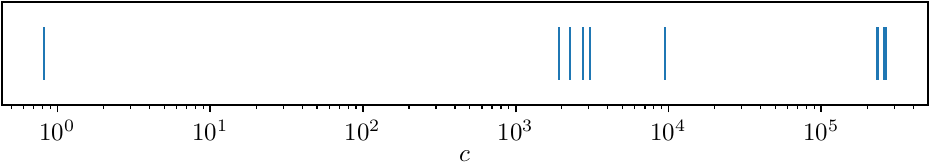}%
	\caption{Similarity measure employed for selecting the new
          eigenvalue pair in each iteration. The logarithmic plot displays the pair-discrepancy values defined in \cref{eq:Toy5DDiscrepancy}. The calculations are performed after the first training step for the model in \cref{fig:Toy5DExampleEnergyKappa}. A noticeable gap is observed between the smallest and second smallest pair-discrepancy values. This indicates that the eigenvalue pair with the lowest $c$ value is highly likely to correspond to the \ac{ep}.} %
	\label{fig:Toy5DDiscrepancy}%
\end{figure} 

Applying this similarity measure to the matrix model in
\cref{fig:Toy5DExampleEnergyKappa} after the first training step,
leads to the $c$ values shown in \cref{fig:Toy5DDiscrepancy}. As
visible, there is indeed a large gap between the smallest and second
smallest $c$ value. This observation strongly suggests that the
eigenvalue pair that has the lowest $c$ value is most likely
associated with the \ac{ep}.
The new pair of eigenvalues is used as additional training point
for the \ac{gpr} model. Repetitive iterations of this procedure yield
a highly precise model, consequently enabling the determination of the
precise location of the \ac{ep}.

For the five-dimensional matrix model
the whole iterative process is depicted in
\cref{fig:Toy5DIterativeProcessModel2}a and
\ref{fig:Toy5DIterativeProcessModel2}b. The initial training set
in this case consists of about 20 points. The \ac{gpr} model slowly
approaches the \ac{ep} until it converges after nine training
steps. The Euclidean distance between the exact \ac{ep} and the last
prediction is $d_\mathrm{e} = 5.526\times 10^{-6}$. In order to explore the
energy plane and thus give more information to the \ac{gpr} model, two
attempts are made to both improve convergence and reduce the number of
diagonalizations. An additional training point is added to the
training set by calculating the distance between the last two
predictions, adding it to the last prediction and computing the exact
eigenvalues at this new $\kappa$ point. If this is executed after the
second iteration, not only the convergence improves significantly, but
also the number of diagonalizations is reduced, both visible in
\cref{fig:Toy5DIterativeProcessModel2}c. The number of training steps
is decreased to three, i.e.\ four diagonalizations, and the Euclidean
distance to the exact \ac{ep} is $d_\mathrm{e} = 1.342\times 10^{-6}$.
Adding the extra training point after the third
iteration can be seen in
\cref{fig:Toy5DIterativeProcessModel2}d. Here, no improved convergence
or reduced number of diagonalizations is observable. Compared to the
original training process, the Euclidean distance is almost identical
with $d_\mathrm{e} = 5.537\times 10^{-6}$.

\subsubsection{Convergence criteria}
\label{sec:convergence}
To terminate the iterative procedure, it is essential to establish a convergence criterion. 
The eigenvalues $\lambda_{\boldsymbol{K}}$ of the covariance matrix $\boldsymbol{K}$ can be calculated for each training iteration. The parameter space (i.e.\ the $\kappa$-space) is the input space of the kernel function and thus of the covariance matrix. As the number of $\kappa$-values increases, the eigenvalues decrease. If there are a lot of training points, especially if they are close together, a drop in the eigenvalues is visible. An interpretation for this drop is that the model has already seen this new training point, thus yielding no significant additional knowledge. 
The utilized data originates from the training process, where an extra training point is incorporated following the second iteration. 
\begin{figure}%
	\centering%
	\includegraphics[width=0.9\textwidth]{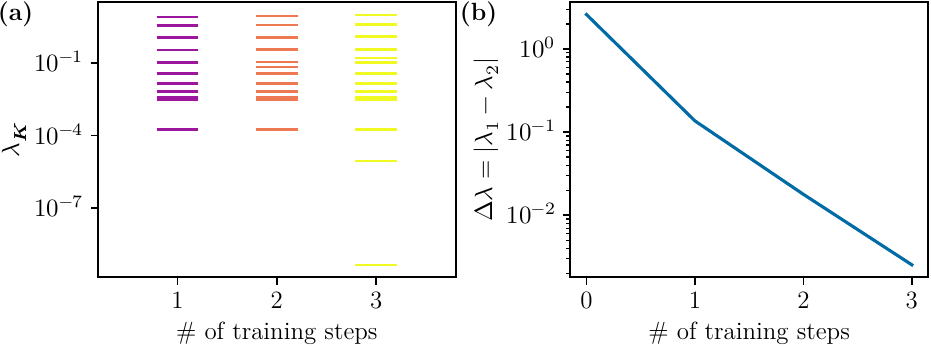}%
	\caption{The two convergence criteria, namely the eigenvalues of the covariance matrix and the eigenvalue difference, are shown for the model system. \textbf{(a)} The eigenvalues $\lambda_{\boldsymbol{K}}$ of the covariance matrix $\boldsymbol{K}$ are depicted for each training step. A drop is visible in the third iteration from order $\mathcal{O}\left(10^{-4}\right)$ to $\mathcal{O}\left(10^{-10}\right)$ compared to the previous step. This indicates an already seen training point that does not provide any new information. A clear deviation to the previous training step is visible. Thus an appropriate threshold value can be defined easily. \textbf{(b)} The eigenvalue difference of the two eigenvalues belonging to the \ac{ep} is plotted over the number of training steps. It decreases strictly monotonically, verifying convergence of the iterative process. Defining a threshold value is not as straightforward as for the kernel eigenvalues, since no clear change is visible between the last two iterations.}%
	\label{fig:Toy5DConvergence}%
\end{figure}%
A drop ($\mathcal{O}\left(10^{-4}\right)$ to $\mathcal{O}\left(10^{-10}\right)$) in the kernel eigenvalues appears in the last training step in \cref{fig:Toy5DConvergence}a. This deviation from the previous training step facilitates the establishment of a threshold. It is important to note that this criterion does not provide insights into the convergence of the \ac{ep} itself.

At the \ac{ep} the difference of the two eigenvalues $\Delta\lambda =
\left|\lambda_1 - \lambda_2\right|$ should be zero due to their
degeneracy. Because of the square root behavior (see
\cref{eq:EPSimpleExampleEigenvalues}) the gradient is infinite at the
\ac{ep}. As a consequence, even slight changes in the $\kappa$-value
can lead to significant variations in the difference between the
eigenvalues. This strong dependency poses a challenge in identifying
an appropriate threshold value for the eigenvalue difference as a
convergence parameter. However, this is a criterion directly related
to a property of an \ac{ep}. In \cref{fig:Toy5DConvergence}b, the
eigenvalue difference $\Delta\lambda$ is plotted as a function of the
number of training steps.
Here, the eigenvalue difference of the initial training set is calculated via
\begin{equation}
	\Delta\lambda_0 = \frac{1}{N} \sum_{i=1}^{N} \left|\lambda_{1,i} - \lambda_{2,i}\right| \,,\label{eq:Toy5DConvergenceEvDiffZerothTrainingStep}
\end{equation}
where $i$ denotes the index of the $i$-th training point $\kappa_i$ on the orbit. As visible, the eigenvalue difference decreases strictly monotonically which verifies the model's convergence towards the \ac{ep}.

\section{Results for exceptional points in \ac{cu2o}}
\label{sec:Results}
In this section, we want to discuss the application of our
\ac{gpr}-based method to a specific physical system and reveal the
existence of \acp{ep} in exciton spectra of cuprous oxide in external
electric and magnetic fields.
In semiconductors like \ac{cu2o}, excitons are
  created when an external photon lifts an electron from the valence
  band to the conduction band, leaving a hole behind. The resulting
  electron-hole pair interacts via the Coulomb interaction, leading to
  bound states that approximately behave like the states known from
  the hydrogen atom, but with modified material parameters.
  Such a hydrogenlike model reproduces the qualitative and some of the quantitative features visible in experiment. When the effects of the crystal structure are additionally taken into account, the resulting model is accurate enough to allow for line-by-line comparisons with experimental spectra~\cite{Schweiner2016Impact,Schweiner2017Magneto}.
  If external electric and magnetic fields are
  added to the system, resonances with finite lifetimes can
  occur. Tuning of the field strengths can then lead to the appearance of \acp{ep}.
More technical details about the system and the numerical computation of the
resonance states are given in Refs.~\cite{Zielinski_2020,Rommel2020Green,Schweiner2017Magneto} and,
for the convenience of the reader, also in \ref{sec:excitonsCu2O}.

\subsection{Application of the \ac{gpr}-based method}

To minimize the number of diagonalizations and thus the computational
cost, the \ac{gpr}-based method is now applied to the Hamiltonian of
\ac{cu2o}.
The Hamiltonian depends on the magnetic and electric field strengths,
$\gamma$ and $f$, respectively.
In analogy to the matrix models described in \cref{sec:MethodsAndMaterials}
the Hamiltonian can be parameterized by a complex parameter $\kappa= \euler^{\imag\phi}$,
which is related to the field strenghts via
\begin{eqnarray}
	\gamma &= \gamma_\mathrm{c} \left(1 + \varrho\, \RE(\kappa)\right)\,, 
	\quad f = f_\mathrm{c} \left(1 + \varrho\, \IM(\kappa)\right)\,, \label{eq:resultsF} 
\end{eqnarray}
with $\varrho = \frac{\Delta\gamma}{\gamma_\mathrm{c}} = \frac{\Delta f}{f_\mathrm{c}}$.
These fields are aligned parallel to the symmetry axis $\left[001\right]$ of \ac{cu2o}.
Here, $\kappa$ is the unit circle in the complex plane and $\varrho$
the so-called relative radius, so the variation of the field
strengths, $\Delta\gamma$ and $\Delta f$ respectively, depends on the
center of the ellipse $\left(\gamma_\mathrm{c}, f_\mathrm{c}\right)$. Due
to this representation, the field strengths on the ellipse can be
converted to their respective points on the unit circle in the complex
plane $\kappa$, which is used to train the \ac{gpr} model.

An ellipse with a radius of $\varrho = 0.06$ is drawn in the field plane, which corresponds to $\Delta\gamma = \SI{77.733}{mT}$ and $\Delta f = \SI{5.1966}{V\,cm^{-1}}$. 
\begin{figure}[t]
	\centering
	\includegraphics[width=0.8\textwidth]{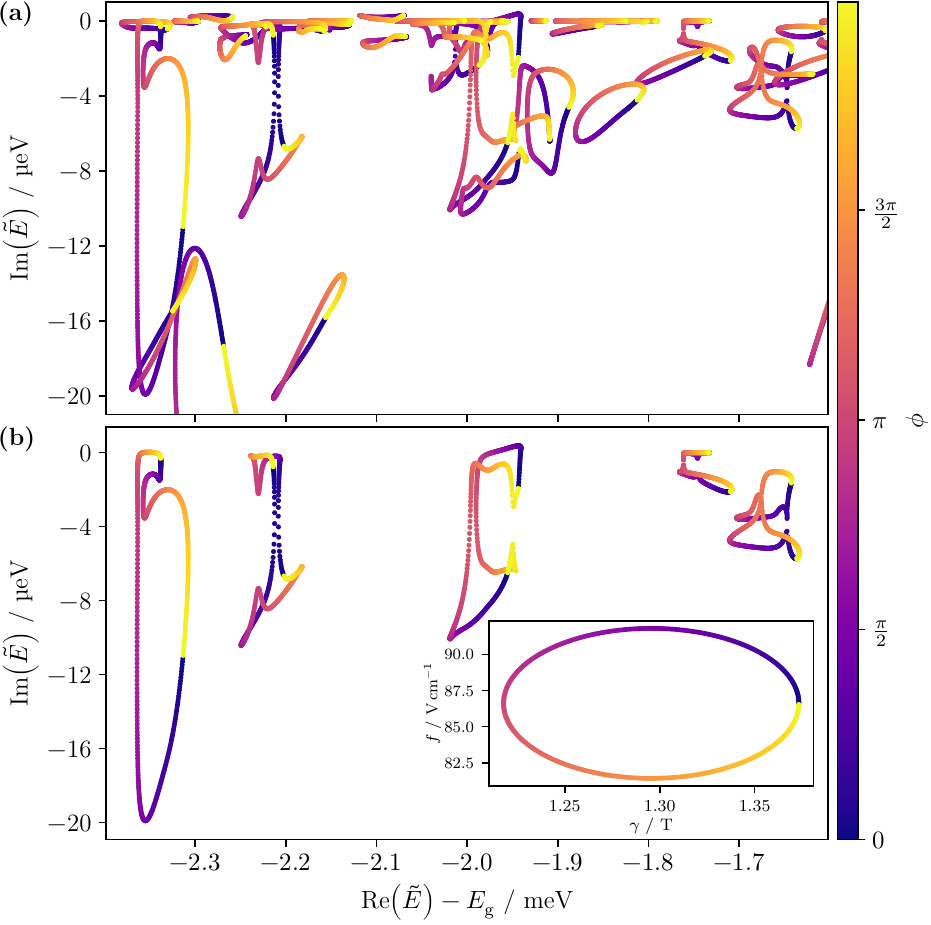}%
	\caption{\textbf{(a)} Drawing an ellipse with a radius of $\varrho = 0.06$, corresponding to $\Delta\gamma = \SI{77.733}{mT}$ and $\Delta f = \SI{5.1966}{V\,cm^{-1}}$, results in the visible resonances for the given energy range. Here, the band gap energy $E_\mathrm{g}$ is subtracted from the real part of the energy. Due to the large radius, there exists a significant overlap with other resonances, resulting in only the leftmost permutation being visually discernible when plotting all eigenvalues for each angle $\phi$. \textbf{(b)} Applying the stepwise grouping algorithm reveals five permutations. Consequently, this algorithm serves as a powerful tool to effectively filter these permutations, enabling the identification of additional \acp{ep} and, subsequently, generating more training data for a single ellipse. This approach effectively reduces the overall computational workload while increasing the number of \acp{ep} discovered.}
\label{fig:ResultsPunkt23BiggestRadiusEnergyMultiplePermutations}
\end{figure}
This ellipse and the observed permutations in the energy plane are
depicted in
\cref{fig:ResultsPunkt23BiggestRadiusEnergyMultiplePermutations}a.
As a result of the large radius, there is a substantial overlap of
resonances belonging to an \ac{ep} and other resonances.
Consequently, when plotting all eigenvalues for each angle
$\phi$ (cf.\
\cref{fig:ResultsPunkt23BiggestRadiusEnergyMultiplePermutations}a),
only the leftmost permutation is visually distinguishable. Applying
the stepwise grouping algorithm yields all permutations shown in
\cref{fig:ResultsPunkt23BiggestRadiusEnergyMultiplePermutations}b.
Therefore,
it functions as a powerful tool to efficiently filter these
permutations, facilitating the discovery of additional \acp{ep}. Thus,
it generates a greater amount of training data for a single
closed loop in the parameter space.
To avoid possible issues with the stepwise grouping
algorithm, the orbit was discretized using 400 points to ensure
that eigenvalues and their respective quantum numbers changed
minimally between steps.
This is of course computationally expensive, but nevertheless the overall computational workload is reduced drastically due to the discovery of additional \acp{ep}.
For the application of \ac{gpr}, the number of points is reduced, in order to prevent numerical inaccuracies leading to negative kernel eigenvalues. The initial training set for all \acp{ep} discussed in this section is thus made up of about {40}-{60} points.

The leftmost permutation in
\cref{fig:ResultsPunkt23BiggestRadiusEnergyMultiplePermutations}b is
taken as initial training set to study the convergence of the method.
\begin{figure}[t]
\centering
\includegraphics[width=0.8\textwidth]{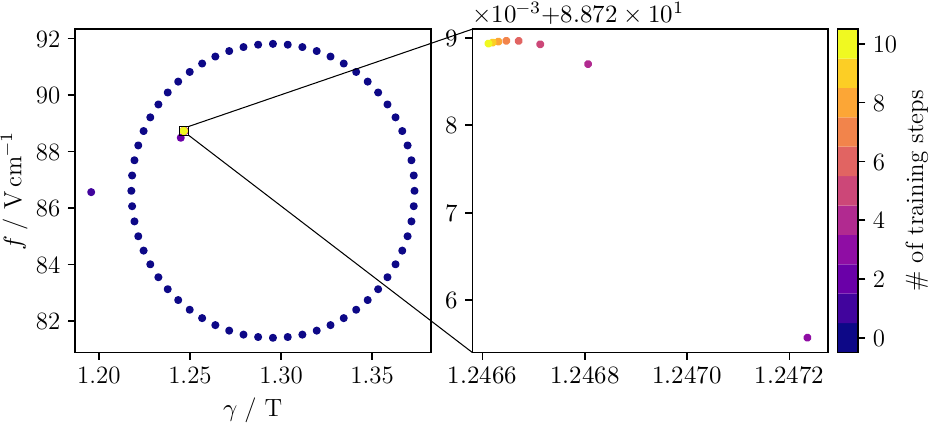}%
\caption{Applying the iterative process to the orbit in \cref{fig:ResultsPunkt23BiggestRadiusEnergyMultiplePermutations} combined with the leftmost permutation as initial training set yields the illustrated prediction of the \ac{ep}. It converges after ten training steps. The Euclidean distance to the prediction of the \ac{ep} in \cref{fig:ResultsPunkt23KappaTrained} is $d_\mathrm{e} = 2.140\times 10^{-4}$, thus the predicted field strengths are identical up to and including the fifth significant digit.}
\label{fig:ResultsPunkt23BiggestRadiusKappaTrained}
\end{figure}
\Cref{fig:ResultsPunkt23BiggestRadiusKappaTrained} visualizes the
result of the iterative process. To ensure its accuracy, a smaller
ellipse around the \ac{ep} is drawn with a relative radius of
$\varrho=0.003$, corresponding to $\Delta\gamma = 
\SI{3.741}{mT}$ and $\Delta f =
\SI{0.266}{V\,cm^{-1}}$.
\begin{figure}[t]
	\centering
	\includegraphics[width=0.8\textwidth]{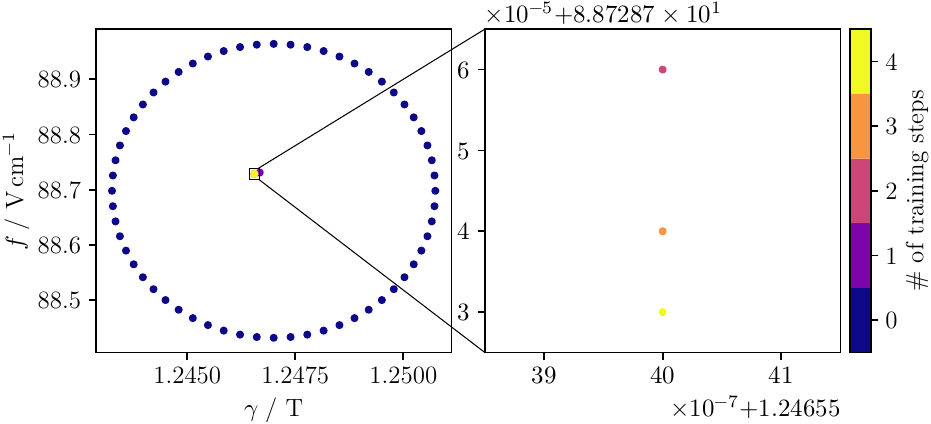}%
	\caption{Applying the iterative process to the small orbit with a relative radius of $\varrho = {0.003}$ yields the illustrated prediction of the \ac{ep}. The iterative process converges already after the fourth training step.}
	\label{fig:ResultsPunkt23KappaTrained}
\end{figure}
\begin{figure}[t]
	\centering
	\includegraphics[width=0.9\textwidth]{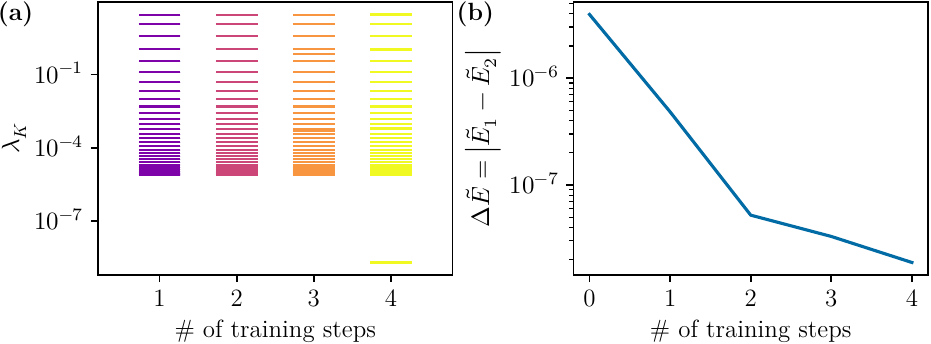}%
	\caption{Convergence of the \ac{ep} by means of the kernel eigenvalues $\lambda_{\boldsymbol{K}}$ and the eigenvalue difference $\Delta\tilde{E}$ of the two eigenvalues belonging to the \ac{ep} for the training process in \cref{fig:ResultsPunkt23KappaTrained}. The significant drop in the kernel eigenvalues from order $10^{-6}$ to $10^{-9}$ is visible. This drop is accompanied by a very small change in the eigenvalue difference, suggesting the convergence of the method.}%
	\label{fig:ResultsPunkt23Punkt28Convergence}
\end{figure}
\begin{table}[t]
\caption{\acp{ep} in \ac{cu2o} found using the \ac{gpr}-based
  method. For each \ac{ep}, the magnetic ($\gamma$) and electric ($f$)
  field strengths as well as the complex energy $\tilde{E}$ are
  given. Here, the band gap energy $E_\mathrm{g}$ is subtracted from the
  real part of the energy.}
\label{tab:ResultsAllEPsFound}
\lineup
\begin{indented}
	\item[]\begin{tabular}{@{}ccccc} 
		\br
		{\ac{ep}} & {$\gamma$ / T} & {$f$ / $\mathrm{V\,cm^{-1}}$} & {$\RE(\tilde{E}) - E_\mathrm{g}$ / meV} & {$\IM(\tilde{E})$ / $\mu$eV} \\
		\mr
		\01 & 1.184658 & \085.348284 & \-1.999440 & \0\-4.103 \\ 
		\02 & 1.218638 & \085.829520 & \-1.761387 & \0\-0.545 \\
		\03 & 1.246554 & \088.728730 & \-2.355771 & \0\-4.942 \\ 
		\04 & 1.252867 & \082.244780 & \-1.675592 & \0\-2.798 \\ 
		\05 & 1.308150 & \087.574636 & \-2.219199 & \0\-4.282 \\
		\06 & 1.338139 & 108.084785 & \-2.331259 & \0\-3.295 \\ 
		\07 & 1.931141 & 117.073400 & \-2.465701 & \0\-2.575 \\ 
		\08 & 1.950923 & 106.762009 & \-2.087715 & \0\-1.249 \\ 
		\09 & 2.049782 & 156.927751 & \-2.363292 & \0\-0.965 \\
		10 & 2.356121 & 143.134723 & \-1.902537 & \-12.337 \\ 
		11 & 2.419380 & 131.213089 & \-1.654835 & \0\-2.685 \\ 
		\br
	\end{tabular}
\end{indented}
\end{table}%

The \ac{gpr}-based method applied to the large ellipse converges after ten training steps and leads to a similar position of the \ac{ep}
\begin{eqnarray}
\gamma_\mathrm{EP,lr} &= \SI{1.246612}{T} \,, \quad 
f_\mathrm{EP,lr} = \SI{88.728936}{V\,cm^{-1}} \label{eq:resultsEPflr}
\end{eqnarray}
as the prediction of the small radius in \cref{fig:ResultsPunkt23KappaTrained}
\begin{eqnarray}
\gamma_\mathrm{EP,sr} &= \SI{1.246554}{T} \,, \quad
f_\mathrm{EP,sr} = \SI{88.728730}{V\,cm^{-1}}\,.
\end{eqnarray}
The indices `lr' and `sr' represent the large and small radius, respectively. Hence, the predicted field strengths are identical at least up to and including the fifth significant digit and the Euclidean distance is only $d_\mathrm{e} = 2.140\times 10^{-4}$. Thus, the iterative process converges and yields a fairly accurate prediction of the position of the \ac{ep} even for a large radius. 

Having a look at the two convergence criteria in \cref{fig:ResultsPunkt23Punkt28Convergence} supports the assumption of a converged training process. 
Not only the drop in the kernel eigenvalues $\lambda_{\boldsymbol{K}}$ from order $10^{-6}$ to $10^{-9}$ for the small ellipse occurs, but also the eigenvalue distance $\Delta\tilde{E}$ of the two eigenvalues associated with the \ac{ep} reduces significantly during the iterative process. 
Furthermore, the drop is accompanied by minimal alteration in the eigenvalue difference, indicating the convergence of the method. Since the exact position of the \ac{ep} is unknown, it is not possible to determine the exactness of the last prediction. However, the deviation only in the sixth significant digit in \cref{fig:ResultsPunkt23KappaTrained} as well as the small eigenvalue difference indicates a very accurate result.

All \acp{ep} found with the \ac{gpr} method are listed in \cref{tab:ResultsAllEPsFound}.
The \ac{gpr} method yields the positions and eigenvalues of the
\acp{ep} with high accuracy, and may serve as a starting point for an
experimental verification of \acp{ep} in cuprous oxide.
However, it should be noted that the
accuracy of the results also depends on the size of the basis set and
convergence parameters for the diagonalization of the cuprous oxide
Hamiltonian \eref{eq:Hamiltonian} with external fields as discussed in \ref{sec:excitonsCu2O}.

\section{Conclusion{s and outlook}}
\label{sec:Conclusion}
In open quantum systems, resonances with a finite lifetime can occur. There are \acp{ep}, where two resonances coalesce, i.e.\ their eigenvalues as well as their eigenvectors degenerate. It is hard to pinpoint them and additionally in systems like \ac{cu2o} the required diagonalizations of the Hamiltonian are computationally expensive. \ac{gpr} serves as a powerful machine learning method to optimize the whole searching process.

The \ac{gpr}-based method was developed gradually by means of matrix
models. An observed permutation of the eigenvalues belonging to the
\ac{ep} was used as initial training set, since it confirms the
existence of an \ac{ep} inside the orbit. In higher-dimensional
systems, it is not straightforward to select the eigenvalues that
perform the permutation due to overlap with other resonances.
A challenge was to extract these eigenvalues, so the stepwise grouping algorithm was developed to solve the initial training set problem. The underlying principle of this approach is to determine the eigenvalue corresponding to the next angle on the orbit for a given resonance. This is accomplished by selecting the eigenvalue that exhibits the closest proximity to the eigenvalue of the current angle, utilizing a suitable distance metric. The grouping algorithm allows filtering the eigenvalues corresponding to the \ac{ep} and hence simplifies obtaining the initial training set. For all parameter values, e.g.\ field strengths, the centroid and the difference of the eigenvalues can be calculated respectively. These are passed to the \ac{gpr} model together with the field strengths. Thus, the latter provides a prediction of these quantities as a function of the field strengths. Due to the degeneracy at the \ac{ep}, the eigenvalue difference has to be zero at this point, which is why a two-dimensional root search applied to the model provides a prediction for the position of the \ac{ep} in the field plane. 
A diagonalization with the predicted field strengths was performed to
obtain the exact eigenvalues. These were added to the training set to
improve the prediction of the model. Hence, an iterative process was
established.
In higher-dimensional systems, diagonalization involves computing more
than just the two eigenvalues associated with the \ac{ep}. Selecting
the correct eigenvalues in each iteration was achieved by introducing
a similarity measure, which compares the model prediction with the
exact eigenvalues.
The smallest value of the similarity measure should correspond to the
eigenvalue pair belonging to the \ac{ep}.
To ensure proper selection, a large gap between the smallest and
second smallest similarity value is desirable, which was indeed observed.
To determine the convergence of this method, a combination of the drop
in the kernel eigenvalues and the difference of the eigenvalues
associated with the \ac{ep} appeared to be a promising convergence
criterion.
Training the \ac{gpr} model and applying the method resulted in a
relatively fast convergence to the \ac{ep} with a high accuracy.
The \ac{gpr}-based method was first developed and tested for a
five-dimensional matrix model, and then successfully applied to find
\acp{ep} in \ac{cu2o}.

In this paper we have focused on the localization of \acp{ep}
consisting of two coalescing resonances, i.e., the case of an EP2.
The \ac{gpr}-based method developed here can certainly be extended and
applied to other more general situations.
For example, the method can, in principle, also be applied to
\acp{dp}, where the eigenvectors corresponding to a degenerate
eigenvalue are orthogonal, contrary to those for \acp{ep}, and the two
eigenvalues show a linear splitting instead of a square root branch
point in the close vicinity of the degeneracy.
In this case, the function $p=(\lambda_1-\lambda_2)^2$ in \cref{eq:EvDiff}
must be replaced with $p=\lambda_1-\lambda_2$ to determine the
position of the \ac{dp}.
However, when encircling a \ac{dp} the two resonances do not show the
characteristic exchange behavior of an \ac{ep}, and therefore a
different strategy than described in \cref{sec:sga} would be necessary
to find candidates for \acp{dp} and initial training sets for the
\ac{gpr}-based method.

While two parameters are sufficient for the coalescence of the two
resonances of an EP2, $2n-2$ parameters must be adjusted for the
coalescence of all resonances of a more general higher-order
EP$n$ with $n\ge 3$~\cite{Heiss2008,Cartarius2009,Hodaei2017}.
(Note that a different number of $(n^2+n-2)/2$ required parameters is
given in reference~\cite{Heiss2008}.)
For the localization of such points, the stepwise grouping algorithm
described in \cref{sec:sga} must be modified to search for the
characteristic exchange behavior of a higher-order EP$n$, and the
functions $p$ and $s$ in \cref{eq:EvDiff,eq:EvSum} need to be adopted
to the $n$th root branch point singularity of the EP$n$, i.e.,
$s=\frac{1}{n}\sum_{i=1}^{n}\lambda_i$ and $p$ must be extended to a set
of complex functions $p_i=(\lambda_{i+1}-\lambda_i)^n$ for $i=1,\dots,n-1$.
The extended \ac{gpr}-based method will then iteratively estimate the
$2n-2$ real parameters of the system to finally achieve $p_i=0$ for
$i=1,\dots,n-1$.
However, it should be noted that although signatures of higher-order
\acp{ep} have been observed, e.g., for the hydrogen atom in external
electric and magnetic fields~\cite{Cartarius2009}, the required number
of parameters for the coalescence of all resonances of an EP$n$ is
often not available in physical systems, such as the hydrogen atom or
cuprous oxide in external fields, which typically depend on a low and
limited number of external parameters.

\section*{Data availability statement}
Within the scope of this research, a Python package
was developed and made publicly available at \url{https://github.com/Search-for-EPs/Search-for-EPs}.
The package is primarily
tailored to address the specific challenges and requirements
encountered in this study, as well as being compatible with other
software utilized in this work. Nonetheless, certain functions within
the package can be applied to other problem domains as well, e.g.\ the
stepwise grouping algorithm.
A documentation for the package is provided at
\url{https://search-for-eps.github.io/Search-for-EPs/}
to facilitate its effective usage.
The data that support the findings of this study are available upon request from the authors.

\ack
We thank Samuel Tovey for useful discussions.
This work was supported by Deutsche Forschungsgemeinschaft (DFG)
through Grant No.\ MA 1639/16-1.

\appendix
\section{Gaussian processes in machine learning}
\label{sec:gprInML}
\acp{gp} are a widely accepted and sophisticated approach used for performing Bayesian non-linear non-parametric regression and classification tasks \cite{snelson_sparse_2005}. They belong to the group of supervised machine learning algorithms \cite{rasmussen_gaussian_2005,mackay_introduction_1998}. 
By incorporating prior knowledge of the function being searched using a kernel, it is possible not only to make predictions, but also to provide model uncertainty for that prediction. 

\subsection{Multivariate normal distribution}

For simplicity, a two-dimensional space is assumed denoted by $x$ and $y$. We generate two
sets of random variables $\boldsymbol{y}_{1,2}$ with component pairs \{$({y}_i)_1$, $({y}_i)_2$\} each sampled from a bi-variate normal distribution given by
\begin{equation}
	\left[\begin{array}{c}({y}_i)_1 \\ ({y}_i)_2\end{array}\right] \sim \mathcal{N}\left(\boldsymbol{\mu}, \boldsymbol{\Sigma}\right) = \mathcal{N}\left(\left[\begin{array}{c}\mu_1 \\ \mu_2\end{array}\right], \left[\begin{array}{cc}\sigma_{11} & \sigma_{12} \\ \sigma_{21} & \sigma_{22}\end{array}\right]\right)\,,
	\label{eq:BND}
\end{equation}
with the mean vector $\boldsymbol{\mu}$ and the covariance
  matrix $\boldsymbol{\Sigma}$.
  Plotting the components of $\boldsymbol{y}_1$ and $\boldsymbol{y}_2$ at $x_1$ and $x_2$ respectively and connecting each value in $\boldsymbol{y}_1$ with the corresponding one in $\boldsymbol{y}_2$ yields a set of linear functions which in general can be used for regression tasks \cite{wang_intuitive_2022}.
The distribution given in \cref{eq:BND} characterizes the joint probability distribution $P(\boldsymbol{y}_1,\boldsymbol{y}_2)$. The covariance matrix $\boldsymbol{\Sigma}$ captures the correlations between $\boldsymbol{y}_1$ and $\boldsymbol{y}_2$ through its off-diagonal elements, $\sigma_{12}$ and $\sigma_{21}$. 
By extending this concept to an infinite number of vectors $\boldsymbol{y}$ at different positions $x$ and connecting values from adjacent vectors, a set of functions can be obtained. To achieve smoother functions, a suitable covariance function $k(x,x')$ can be employed to account for correlations among nearby values. Thus, an infinite-dimensional Gaussian can be used to produce a continuous function space.

\subsection{Gaussian processes}

A \ac{gp} is defined as a set of random variables where the joint distribution of any finite subset follows a Gaussian distribution, therefore it describes a distribution over functions \cite{rasmussen_gaussian_2005}. For a given function, $f(\boldsymbol{x})$, the mean $m(\boldsymbol{x})$ and covariance function $k(\boldsymbol{x},\boldsymbol{x}')$ are defined as 
\begin{equation}
	m(\boldsymbol{x}) = \mathbb{E}[f(\boldsymbol{x})] \,, \label{eq:GPMeanFunction}%
\end{equation}%
\begin{equation}%
	k(\boldsymbol{x}, \boldsymbol{x}') = \mathbb{E}[(f(\boldsymbol{x}) - m(\boldsymbol{x}))(f(\boldsymbol{x}')- m (\boldsymbol{x}'))]\,, \label{eq:GPCovarianceFunction}
\end{equation}
which fully specify a \ac{gp} via
\begin{equation}
	f(\boldsymbol{x}) \sim \mathcal{GP}(m(\boldsymbol{x}), k(\boldsymbol{x}, \boldsymbol{x}')) \,.
\end{equation}
The vector $\boldsymbol{x} \in X \subset \mathbb{R}^D$ can be $D$-dimensional and belongs to the input set $X$ which contains all possible inputs. At a particular position $\boldsymbol{x}$ the function value $f(\boldsymbol{x})$ is expressed by the random variables.

\subsection{Prior and posterior of a Gaussian process}

A linear regression model
\begin{equation}
	f(\boldsymbol{x}) = \boldsymbol{\phi}(\boldsymbol{x})^\top \boldsymbol{w}
	\label{eq:GPLinearRegression}
\end{equation}
can provide a straightforward instance of a \ac{gp}. Specifying a prior over the parameters --- usually a Gaussian $\boldsymbol{w} \sim \mathcal{GP}(\boldsymbol{0}, \Sigma_p)$ with zero mean and covariance matrix $\Sigma_p$ --- is necessary to encapsulate assumptions regarding these parameters prior to inspecting the observed data.
\begin{figure}
	\centering
	\includegraphics[width=0.8\textwidth]{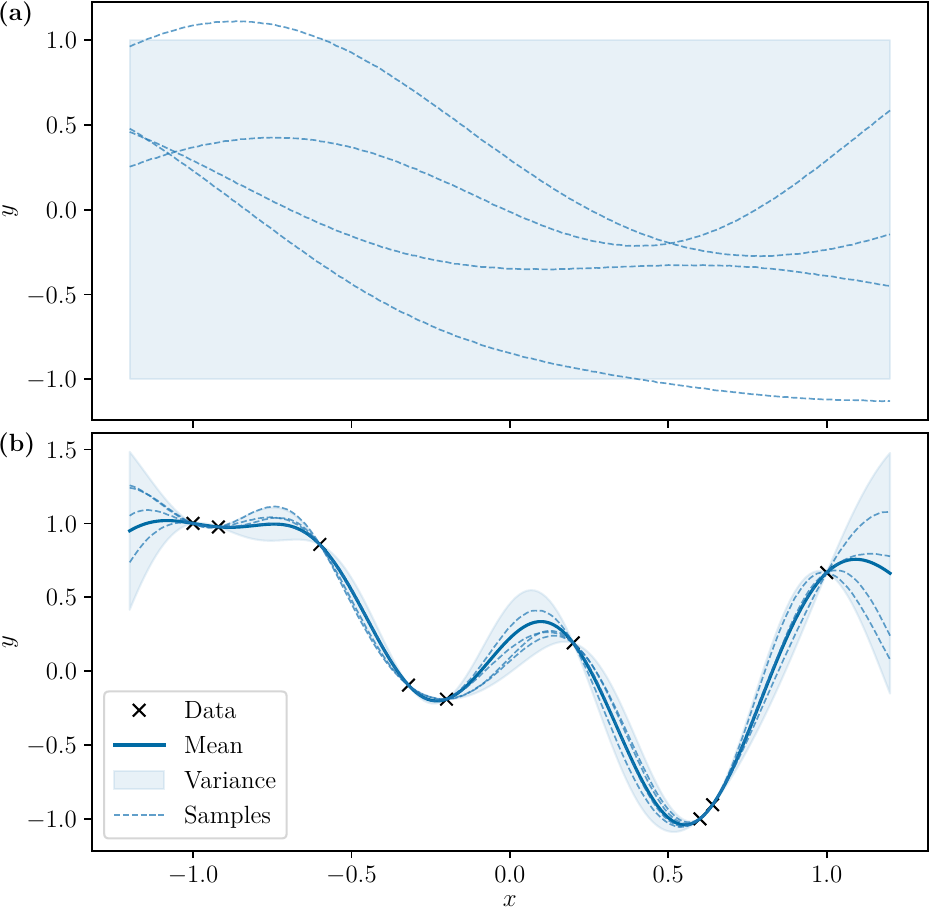}%
	\caption{\textbf{(a)} A \ac{gp} prior distribution is visualized, including four sample functions which are randomly chosen from this prior. Due to a large number of given input points, they are displayed as dashed lines rather than individual points, which would be more accurate. \textbf{(b)} The posterior distribution obtained after fitting the model to the observed, noise free, data points, marked as black crosses. Again, four sample functions randomly selected from the posterior are shown as dashed lines. Taking the mean of all possible functions that fit the given data points results in the solid blue line, which is used to make predictions. The shaded region in both plots marks the $\SI{95}{\%}$ confidence interval calculated from the variance of the model. Adapted from \cite{rasmussen_gaussian_2005}.} %
	\label{fig:GPRPriorPosterior}
\end{figure}
A \ac{gp} prior distribution and some sample functions randomly selected from it are illustrated in \cref{fig:GPRPriorPosterior}a.
Inserting \cref{eq:GPLinearRegression} and the above assumption about the weights $\boldsymbol{w}$ into \cref{eq:GPMeanFunction,eq:GPCovarianceFunction} results in 
\begin{eqnarray}
	m(\boldsymbol{x}) &= \boldsymbol{\phi}(\boldsymbol{x})^\top \mathbb{E}[\boldsymbol{w}] = 0 \,, \\
	k(\boldsymbol{x}, \boldsymbol{x}') &= \boldsymbol{\phi}(\boldsymbol{x})^\top \mathbb{E}[\boldsymbol{w}\boldsymbol{w}^\top] \boldsymbol{\phi}(\boldsymbol{x}') = \boldsymbol{\phi}(\boldsymbol{x})^\top \Sigma_p \boldsymbol{\phi}(\boldsymbol{x}')
\end{eqnarray}
for the mean and covariance function, respectively. 

It is essential to make a clear distinction between the prior and the posterior distributions.
The posterior distribution over the weights $\boldsymbol{w}$ is used to make predictions as depicted in \cref{fig:GPRPriorPosterior}b. It contains all known information, i.e.\ it is obtained after fitting the model to the observed data points.

\subsection{Covariance and kernel function}

The covariance function, also referred to as kernel function or simply kernel, determines the level of correlation between two random variables \cite{rasmussen_gaussian_2005}. A popular kernel is the \ac{rbf}, which is given by
\begin{equation}
	\mathrm{cov}\left(f(\boldsymbol{x}_p), f(\boldsymbol{x}_q)\right) = k(\boldsymbol{x}_p, \boldsymbol{x}_q) = \euler^{-\frac{1}{2} \left|\boldsymbol{x}_p - \boldsymbol{x}_q\right|^2}
\end{equation}
and thus only depends on the distance between the inputs $\boldsymbol{x}_p$ and $\boldsymbol{x}_q$. 

Mercer's theorem \cite{mercer_xvi_1909,sun_mercer_2005,rasmussen_gaussian_2005} states that any positive definite kernel function can be represented using a possibly infinite number of basis functions, e.g.\ the linear combination of an infinite number of basis functions with Gaussian shapes results in the \ac{rbf} covariance function. Hence, defining the kernel function corresponds to a distribution over possible functions.

The kernel has to be selected according to the problem, since the smoothness of the basis functions, and thus that of the model, depends on it. 
For instance the \ac{rbf} is infinitely mean-square differentiable and thus has smooth basis functions (cf.\ \cref{fig:GPRPriorPosterior}). To achieve better results, the kernel usually contains free positive parameters that can be varied, so called hyperparameters $\Theta$. These are the function variance $\sigma_f^2$ and the characteristic length-scale $l$. If, for example, experimental data with noise is used for training, there is an additional hyperparameter, the noise variance $\sigma_n^2$. An alternative to the \ac{rbf} kernel function is the Matérn class of kernels
\begin{equation}
	k_\mathrm{Mat\acute{e}rn}(r) = \sigma_f^2\frac{2^{1-\nu}}{\Gamma(\nu)} \left(\sqrt{2\nu}r\right)^\nu K_\nu \left(\sqrt{2\nu}r\right) \,,~
	r = \frac{\left|\boldsymbol{x}_p - \boldsymbol{x}_q\right|}{l}\,,
	\label{eq:GPRMaternClassKernel}
\end{equation}
where $\nu$ is a positive parameter and $K_\nu$ a modified Bessel function \cite{rasmussen_gaussian_2005}. The Euclidean distance of $\boldsymbol{x}_p$ and $\boldsymbol{x}_q$ is scaled by the characteristic length-scale $l$. For $\nu \to \infty$ the \ac{rbf} kernel is obtained. Inserting half-integer values for $\nu$ in \cref{eq:GPRMaternClassKernel} results in quite simple equations. In particular, $\nu = \frac{5}{2}$ is widely used in machine learning. It is twice mean-square differentiable and has the form
\begin{equation}
	k(\boldsymbol{r}) = \sigma_f^2 \left(1 + \sqrt{5}r + \frac{5}{3} r^2\right) \euler^{-\sqrt{5}r} + \sigma_n^2\delta_{pq} 
	\label{eq:Matern5_2}
\end{equation}
considering all hyperparameters, where $\delta_{pq}$ is the Kronecker delta. If the inputs are multivariate, usually a separate length-scale $l$ is introduced for each dimension. It is crucial to highlight that the selection of the kernel not only influences the accuracy of the predictions but also affects the quality of the model uncertainty.

\subsection{Gaussian process regression}
\label{sec:GPRTraining}

In order to make predictions, the posterior is needed, which contains the knowledge about the observed data used for training. For the sake of generality, noisy data
\begin{equation}
	y = f(\boldsymbol{x}) + \epsilon
\end{equation}
is assumed with independent Gaussian noise $\epsilon$, which has the variance $\sigma_n^2$. This changes the prior from \cref{eq:GPCovarianceFunction} analogously to \cref{eq:Matern5_2}, where the variance $\sigma_n^2$ is simply added to the kernel with a Kronecker delta. The points used for training are $\left\{(\boldsymbol{x}_i, y_i) | i = 1, \ldots, n\right\}$.
According to the prior, the joint distribution of the training and the test outputs, $\boldsymbol{y}$ and $\boldsymbol{f}_*$, is \cite{rasmussen_gaussian_2005}
\begin{equation}
	\left[\begin{array}{c}\boldsymbol{y} \\ \boldsymbol{f}_*\end{array}\right] \sim \mathcal{GP}\left(0, \left[\begin{array}{cc}\boldsymbol{K}(X,X) + \sigma_n^2 \mathbf{1} & \boldsymbol{K}(X,X_*) \\ \boldsymbol{K}(X_*, X) & \boldsymbol{K}(X_*,X_*)\end{array}\right]\right) \,.
\end{equation}
Here, $X_*$ is the set of new input points at which the predictions $\boldsymbol{f}_*$ of \ac{gp} are sought. Assuming there are $n_*$ number of test points, then $\boldsymbol{K}(X,X_*)$ is an $n \times n_*$ matrix containing all pairwise covariances between the training and test points. The same applies for $\boldsymbol{K}(X,X)$, $\boldsymbol{K}(X_*,X_*)$ and $\boldsymbol{K}(X_*,X)$. Considering only the functions which traverse through the training points results in the posterior distribution. This is illustrated in \cref{fig:GPRPriorPosterior}. For this purpose, the joint Gaussian prior distribution is conditioned on the training points, leading to
\begin{eqnarray}
	\boldsymbol{f}_*|X, \boldsymbol{y}, X_* &\sim \mathcal{GP}\left(\bar{\boldsymbol{f}}_*, \mathrm{cov}(\boldsymbol{f}_*)\right)\,, \label{eq:GPRPrediction}
\end{eqnarray}
with
\begin{equation}
	\fl\boldsymbol{\mu} (X_* | X, \boldsymbol{y}) = \bar{\boldsymbol{f}}_* = \mathbb{E}[\boldsymbol{f}_*|X,\boldsymbol{y}, X_*] = \boldsymbol{K}(X_*,X)[\boldsymbol{K}(X,X) + \sigma_n^2 \mathbf{1}]^{-1}\boldsymbol{y} \,, \label{eq:GPRMeanFunction}
\end{equation}%
\begin{equation}%
	\fl\boldsymbol{\sigma}^2 (X_* | X) = \mathrm{cov}(\boldsymbol{f}_*) = \boldsymbol{K}(X_*,X_*) - \boldsymbol{K}(X_*,X)[\boldsymbol{K}(X,X) + \sigma_n^2 \mathbf{1}]^{-1}\boldsymbol{K}(X,X_*) \,. \label{eq:GPRCovarianceFunction}
\end{equation}
These are the key predictive equations for \ac{gpr} \cite{rasmussen_gaussian_2005}, resulting in the posterior distribution, with the mean function $\boldsymbol{\mu}$ $(\mu_i = m(\boldsymbol{x}_i))$ in \cref{eq:GPRMeanFunction} and the covariance function $\boldsymbol{\sigma}^2$ in \cref{eq:GPRCovarianceFunction}, that can be used to predict $\boldsymbol{f}_*$ at the respective input points $X_*$, as shown in \cref{fig:GPRPriorPosterior}b. 

To optimize the hyperparameters $\Theta$, the \ac{lml} is required, which is defined as
\begin{equation}
	\fl\log p(\boldsymbol{y} | X) = -\frac{1}{2} \boldsymbol{y}^\top (\boldsymbol{K}(X,X) + \sigma_n^2 \mathbf{1})^{-1}\boldsymbol{y} - \frac{1}{2} \log\left|\boldsymbol{K}(X,X) + \sigma_n^2 \mathbf{1}\right| - \frac{n}{2} \log 2\pi \,. \label{eq:LogMarginalLikelihood}
\end{equation}
Note that the scalar $p$ here denotes the marginal likelihood, and should not be confused with the difference of eigenvalues in~\cref{eq:EvDiff}.
A short characteristic length-scale that might improve the data fit is not preferred by the \ac{lml}. Rather, maximizing the \ac{lml} (maximum likelihood estimation) leads to a value which increases the likelihood that the training data would be generated by the distribution over functions \cite{lotfi_bayesian_2022}. This technique is particularly powerful because the \ac{lml} is differentiable with respect to the hyperparameters $\Theta_i$ and utilizes only the available training data.
There are different Python packages to perform \ac{gpr}. The one used
in this work is GPflow~\cite{gpflow}.

\section{Excitons in cuprous oxide}
\label{sec:excitonsCu2O}
We are interested in excitons in cuprous oxide under the influence of
an external magnetic and electric field.
We thus briefly recapitulate the description of the system we want to
use in this paper. 

The field-free spectrum of yellow excitons in \ac{cu2o} is dominated by the odd parity $P$ states with an additional fine structure splitting of other odd parity states, like $F$ states~\cite{Kazimierczuk2014,Thewes2015}. For their accurate modelling, one needs to go beyond the qualitative hydrogen-like description and consider the non-parabolicity of the valence band $H_{\rm b}(\boldsymbol{p})$. Since we also want to apply external fields, especially an external electric field, mixing between the odd- and even-parity states needs to be taken into account. Due to their small extention, $S$ states probe features of the crystal on short length scales, requiring the inclusion of the central-cell corrections $H_{\mathrm{CCC}}$.
With these, the total field-free Hamiltonian
is given by~\cite{Rommel2020Green,Schweiner2016Impact,Schoene2016,Luttinger1956,Schweiner2017Even}
\begin{equation}
	H = E_{\rm g}+\frac{\gamma_1'}{2m_{0}}\boldsymbol{p}^2+H_{\rm b}(\boldsymbol{p})
	-\frac{e^2}{4\pi\varepsilon_0\varepsilon|\boldsymbol{r}|}\ + H_{\mathrm{CCC}}\, ,
	\label{eq:Hamiltonian}
\end{equation}
with the corrections due to the non-parabolicity of the valence band
\begin{eqnarray}
	H_{\rm b}(\boldsymbol{p}) &= H_{\rm SO}+\frac{1}{2\hbar^2m_0}\nonumber
	\{4\hbar^2\gamma_2\boldsymbol{p}^2\\[1ex]
	&+2(\eta_1+2\eta_2)\boldsymbol{p}^2(\boldsymbol{I}\cdot\boldsymbol{S}_{\rm h})\nonumber \\[1ex]
	&-6\gamma_2(p^2_{1}\boldsymbol{I}^2_1+{\rm c.p.})
	-12\eta_2(p^2_{1}\boldsymbol{I}_1\boldsymbol{S}_{\rm h1}+{\rm c.p.})\nonumber \\[1ex]
	&-12\gamma_3(\{p_{1},p_{2}\}\{\boldsymbol{I}_1,\boldsymbol{I}_2\}+{\rm c.p.})\nonumber \\[1ex]
	 &-12\eta_3(\{p_{1},p_{2}\}(\boldsymbol{I}_1\boldsymbol{S}_{\rm h2}
	+\boldsymbol{I}_2\boldsymbol{S}_{\rm h1})+{\rm c.p.})\} \,.
	\label{eq:HoleKinetic}
\end{eqnarray}
These terms require the introduction of additional degrees of freedom for the hole, i.e.,
the hole spin $\boldsymbol{S}_\mathrm{h}$ and the quasispin $\boldsymbol{I}$. The spin-orbit
interaction
\begin{equation}
	H_{\rm SO}=\frac{2}{3}\Delta
	\left(1+\frac{1}{\hbar^2}\boldsymbol{I}\cdot\boldsymbol{S}_{\rm h}\right)\,,
	\label{eq:SOCoupling}
\end{equation}
splits the valence bands into the uppermost $\Gamma_7^+$ and the lower
$\Gamma_8^+$ bands. The Hamiltonian~\eref{eq:Hamiltonian} is
expressed in center-of-mass coordinates
\cite{Schweiner2016Impact,Schmelcher1992},
\begin{eqnarray}
	\boldsymbol{r}&= \boldsymbol{r}_{\rm e}-\boldsymbol{r}_{\rm h}\, ,\quad
	\boldsymbol{R}=\frac{m_{\rm h}\boldsymbol{r}_{\rm h}+m_{\rm e}\boldsymbol{r}_{\rm e}}{m_{\rm h}+m_{\rm e}}\, ,\nonumber\\
	\boldsymbol{P}&=\hbar\boldsymbol{K}= \boldsymbol{p}_{\rm e}+\boldsymbol{p}_{\rm h}\, ,\quad
	\boldsymbol{p}=\hbar\boldsymbol{k}=\frac{m_{\rm h}\boldsymbol{p}_{\rm e}-m_{\rm e}\boldsymbol{p}_{\rm h}}{m_{\rm h}+m_{\rm e}} \, ,
	\label{eq:COMCoordinates}
\end{eqnarray}
with the total momentum set to zero. 
For the accurate description of the even parity states, the central-cell corrections are given by
\begin{equation}
	H_{\mathrm{CCC}} = V^{\mathrm{PB}} + V_d + H_{\mathrm{exch}}\,,
	\label{eq:CCC}
\end{equation}
Here,
\begin{eqnarray}
	V^{\mathrm{PB}}\!\left(r\right)  = 
	&-\frac{e^{2}}{4\pi\varepsilon_{0}r\varepsilon_{\mathrm{s}1}} \nonumber
	-\frac{e^{2}}{4\pi\varepsilon_{0}r\varepsilon_{1}^{*}}\left(\frac{m_\mathrm{h}}{m_\mathrm{h}-m_{\mathrm{e}}}\euler^{-r/\rho_{\mathrm{h}1}}-\frac{m_{\mathrm{e}}}{m_\mathrm{h}-m_{\mathrm{e}}}\euler^{-r/\rho_{\mathrm{e}1}}\right) \\[1ex] 
	&-\frac{e^{2}}{4\pi\varepsilon_{0}r\varepsilon_{2}^{*}}\left(\frac{m_\mathrm{h}}{m_\mathrm{h}-m_{\mathrm{e}}}\euler^{-r/\rho_{\mathrm{h}2}}-\frac{m_{\mathrm{e}}}{m_\mathrm{h}-m_{\mathrm{e}}}\euler^{-r/\rho_{\mathrm{e}2}}\right)  \label{eq:PBPotential}
\end{eqnarray}
is the Pollmann-Büttner potential,
\begin{equation}
	H_{\mathrm{exch}}= J_0\left(\frac{1}{4} - \frac{1}{\hbar^2}\boldsymbol{S}_\mathrm{e}\cdot\boldsymbol{S}_\mathrm{h}\right) V_\mathrm{uc} \delta(\boldsymbol{r})
	\label{eq:ExchangeInteraction}
\end{equation}
is the exchange interaction \cite{Uihlein1981},
and
\begin{equation}
	V_d = - V_0 V_\mathrm{uc} \delta(\boldsymbol{r})
	\label{eq:ShortDistanceCorrection}
\end{equation}
is an additional short distance correction \cite{Kavoulakis1997}.
The polaron radii are given by
\begin{equation}
	\rho_{\mathrm{e/h},i} = \sqrt{\frac{\hbar}{2m_\mathrm{e/h}\omega_{\mathrm{LO},i}}} 
\end{equation}
with the energies $\hbar\omega_{\mathrm{LO}i}$ of the longitudinal $\Gamma_4^-$ phonons,
and the values
\begin{equation}
	\frac{1}{\epsilon^\ast_i} = \frac{1}{\epsilon_{\mathrm{b}i}} - \frac{1}{\epsilon_{\mathrm{s}i}}\,.
\end{equation}
$V_\mathrm{uc} = a^3$ is the volume of the unit cell.

For the investigation of \acp{ep}, we need to apply at least two
tunable parameters to the Hamiltonian. For this, we use
parallel magnetic and electric fields.
The former can be added via the minimal substitution
$\boldsymbol{p}_\mathrm{e}$ $\rightarrow$ $\boldsymbol{p}_\mathrm{e} + e \boldsymbol{A}$, 
$\boldsymbol{p}_\mathrm{h}$ $\rightarrow$ $\boldsymbol{p}_\mathrm{h} - e \boldsymbol{A}$
and the supplementation of the interaction term of the magnetic field and the spin degrees of freedom~\cite{Schweiner2017Magneto,Luttinger1956},
\begin{equation}
	H_{B}=\mu_{\mathrm{B}}\left[g_{c}\boldsymbol{S}_{\mathrm{e}}+\left(3\kappa+g_{s}/2\right)\boldsymbol{I}-g_{s}\boldsymbol{S}_{\mathrm{h}}\right]\cdot\boldsymbol{B}/\hbar\,,
\end{equation}
with the Bohr magneton $\mu_{\mathrm{B}}$, the $g$-factor of the electron $g_c$ and the hole $g_s \approx 2$, and the fourth Luttinger parameter $\kappa$.

For the electric field, we add the term~\cite{knox1963excitons}
\begin{equation}
	H_\mathcal{F} = e \boldsymbol{\mathcal{F}} \cdot (\boldsymbol{r}_\mathrm{e} - \boldsymbol{r}_\mathrm{h})= e \boldsymbol{\mathcal{F}} \cdot \boldsymbol{r}\,.
\end{equation}
The relevant matrix element of the coordinate vector $\boldsymbol{r}$ is given in \ref{sec:matElec}.
All material parameters are listed in \cref{tab:MaterialParameters}.

\begin{table}[b]
	\renewcommand{\arraystretch}{1.2}
	\caption{Material parameters of Cu$_2$O used in the calculations.}
	\begin{indented}
	\item[]\begin{tabular}{@{}llc}
		\br
		Energy gap  & $E_{\rm g}=\SI{2.17208}{eV}$ & \cite{Kazimierczuk2014}\\
		Spin-orbit coupling       & $\Delta=\SI{0.131}{eV}$ &\cite{Schoene2016}\\
		Effective electron mass  & $m_{\rm e}=0.99\,m_0$ & \cite{Hodby1976} \\
		Effective hole mass  & $m_{\rm h}=0.58\,m_0$ & \cite{Hodby1976} \\
		Luttinger parameters & $\gamma_1=1.76$&\cite{Schoene2016}\\
		~~& $\gamma_2=0.7532$&\cite{Schoene2016}\\
		~~& $\gamma_3=-0.3668$&\cite{Schoene2016}\\
		~~& $\eta_1=-0.020$&\cite{Schoene2016}\\
		~~& $\eta_2=-0.0037$&\cite{Schoene2016}\\
		~~& $\eta_3=-0.0337$&\cite{Schoene2016}\\
		Exchange interaction & $J_0 = \SI{0.792}{eV}$ & \cite{Schweiner2017Even} \\
		Short distance correction & $V_0 = \SI{0.539}{eV}$ & \cite{Schweiner2017Even}\\
		Lattice constant & $a=\SI{0.42696}{nm}$ & \cite{Swanson1953}\tabularnewline
		Dielectric constants & $\varepsilon_{\mathrm{s}1}=\varepsilon=7.5$ & \cite{LandoltBornstein1998DielectricConstant}\tabularnewline
		& $\varepsilon_{\mathrm{b}1}=\varepsilon_{\mathrm{s}2}=7.11$ & \cite{LandoltBornstein1998DielectricConstant}\tabularnewline
		& $\varepsilon_{\mathrm{b}2}=6.46$ & \cite{LandoltBornstein1998DielectricConstant}\tabularnewline
		Energy of $\Gamma_{4}^{-}$-LO phonons & $\hbar\omega_{\mathrm{LO1}}=\SI{18.7}{meV}$ & \cite{Kavoulakis1997}\tabularnewline
		& $\hbar\omega_{\mathrm{LO2}}=\SI{87}{meV}$ & \cite{Kavoulakis1997}\tabularnewline
		\br
	\end{tabular}
	\label{tab:MaterialParameters}
	\end{indented}
\end{table}

\subsection{Complex coordinate-rotation and numerical solution}
\label{sec:NumSol}
For the treatment of complex resonances with finite linewidths, we
use the complex-coordinate-rotation method~\cite{Reinhardt1982,Ho1983,Moiseyev1998,Zielinski_2020}.
For the numerical solution of the Schrödinger equation we proceed analogously to Ref.~\cite{Rommel2021Interseries},
using the Coulomb-Sturmian functions~\cite{Caprio2012} to obtain a
generalized eigenvalue problem with dimension up to $\sim 10000 \times 10000$.
The solution is calculated numerically, using a suitable ARPACK~\cite{arpackuserguide} routine first for the Hamiltonian
without the singular delta terms in \cref{eq:ExchangeInteraction,eq:ShortDistanceCorrection}.
These are then included for a second diagonalization using a LAPACK~\cite{lapackuserguide3}
routine with only a few converged eigenvalues from the first diagonlization.
Note that the \acp{ep} found by the \ac{gpr}-based method are in parameter regimes where the numerical convergence of the exciton Hamiltonian is hard to ensure. Whether the computed \acp{ep} are converged accurately enough to be the basis for an experimental exploration is subject of further investigation.

\subsection{Matrix elements for the electric field}
\label{sec:matElec}
The Hamiltonian of the electric field requires the matrix element of the coordinates $x$, $y$, and $z$.
We express these through the spherical tensors $R^{(1)}_q$ with $q=0,\pm1$,
\begin{equation}
	R^{(1)}_{\pm 1} = \mp \frac{1}{\sqrt{2}} (x \pm \mathrm{i} y), \quad  R^{(1)}_0 = z\,.
\end{equation}
Using the same basis set as in Ref.~\cite{Schweiner2016Impact}, we get
\begin{eqnarray}
\fl\left\langle \Pi'\left|R^{(1)}_q \right|\Pi\right\rangle
=\delta_{J',J}(-1)^{F'_\mathrm{t}+F_\mathrm{t}-M'_{F_\mathrm{t}}+F'+F+L'+J+\frac{1}{2}}\,
\sqrt{(2F_\mathrm{t}+1)(2F'_\mathrm{t}+1)(2F+1)(2F'+1)}\nonumber\\[1ex]
\fl\times\left(\begin{array}{ccc}
F'_\mathrm{t} & 1 & F_\mathrm{t}\\
-M'_{F_\mathrm{t}} & q & M_{F_\mathrm{t}}
\end{array}\right)%
\left\{\begin{array}{ccc}
F' & F'_\mathrm{t} & \frac{1}{2}\\
F_\mathrm{t} & F & 1
\end{array}\right\}%
\left\{\begin{array}{ccc}
L' & F' & J\\
F & L & 1
\end{array}\right\}
\left\langle N' L' \left|\left| R^{(1)} \right|\right| N L\right\rangle\,,
\end{eqnarray}
with the reduced matrix element
\begin{eqnarray}
  &\left\langle N'\, L'\left\Vert R^{(1)}\right\Vert N\, L\right\rangle\nonumber\\[1ex]
  &= \: \frac{\delta_{L',L+1}}{L+1}\left[\frac{\left(2L+3\right)\left(2L+2\right)\left(2L+1\right)}{2}\right]^{\frac{1}{2}}\left[\sum_{j=-3}^{1}\frac{\left(LN_{1}\right)_{NL0}^{j\,1}}{(N+L+j+2)}\,\delta_{N',N+j}\right]\nonumber \\[1ex]
 &- \: \frac{\delta_{L',L-1}}{L}\left[\frac{\left(2L+1\right)\left(2L\right)\left(2L-1\right)}{2}\right]^{\frac{1}{2}}\left[\sum_{j=-1}^{3}\frac{\left(LN_{1}\right)_{NL0}^{j\,-1}}{(N+L+j)}\,\delta_{N',N+j}\right]\,.
\end{eqnarray}
The coefficients $\left(LN_{1}\right)_{NLM}^{j\,k}$ are given in the appendix of Ref.~\cite{Schweiner2016Impact}.

\section*{References}
\bibliographystyle{iopart-num}
\providecommand{\newblock}{}

\end{document}